\documentclass[preprint]{imsart}

\usepackage{amsthm,amsmath,natbib}

\startlocaldefs
\usepackage{algorithm}
\usepackage{algorithmic,amssymb}
\usepackage{graphicx,xcolor}
\usepackage{subfigure}

\newcommand{\argmax}{\mathop{\mathrm{argmax}}}
\newcommand{\argmin}{\mathop{\mathrm{argmin}}}

\newcommand{\maximize}{\mathop{\mathrm{maximize}}}

\newcommand{\real}{\mathbb{R}}

\newtheorem{theorem}{Theorem}
\newtheorem{lemma}[theorem]{Lemma}

\newcommand{\norm}[1]{\lVert #1 \rVert}
\newcommand{\innerp}[2]{\langle #1 , #2 \rangle}

\newcommand{\pen}{\mathcal{P}}
\newcommand{\vecsp}{\mathcal{L}}
\newcommand{\gstar}{g^*}

\endlocaldefs

\begin{document}

\begin{frontmatter}

\title{A significance test for forward stepwise model selection}

\runtitle{Forward stepwise significance test}

\begin{aug}
\author{\fnms{Joshua R.} \snm{Loftus}\corref{}\ead[label=e1]{joftius@stanford.edu}}
\and
\author{\fnms{Jonathan E.} \snm{Taylor}\ead[label=e2]{jonathan.taylor@stanford.edu}\thanksref{t2}}
\thankstext{t2}{Supported in part by NSF grant DMS 1208857 and AFOSR grant 113039.}

\affiliation{Stanford University}

\address{Department of Statistics\\  Stanford University\\ Sequoia Hall \\390 Serra Mall\\ Stanford, CA 94305, U.S.A.\\ }
\printead{e1,e2}

\runauthor{J. R. Loftus et al.}

\end{aug}

\begin{abstract}
  We apply the methods developed by \cite{significance:lasso} and
  \cite{tests:adaptive} on significance tests for penalized
  regression to forward stepwise model selection. A general framework
  for selection procedures described by quadratic inequalities includes
  a variant of forward stepwise with grouped variables, allowing us to
  handle categorical variables and factor models. We provide an algorithm
  to compute a new statistic with an exact null distribution conditional
  on the outcome of the model selection procedure. This new
  statistic, which we denote $T\chi$,
  has a truncated $\chi$ distribution under the global null.
  We apply this test in forward stepwise iteratively on the residual after
  each step. The resulting method
  has the computational strengths of stepwise selection and addresses
  the problem of invalid test statistics due to model selection.
  We illustrate the flexibility of this method by applying it to
  several specialized applications of forward stepwise including a
  hierarchical interactions model and a recently described additive model
  that adaptively chooses between linear and nonlinear effects for
  each variable. 
\end{abstract}



\end{frontmatter}

\section{Introduction}
\label{sec:intro}

Consider the regression setting with a single response variable $Y$
and collection of predictor or covariate variables denoted $\mathcal X$.
One often wishes to choose a subset of variables $X \subset \mathcal X$
for modeling the response, assuming the remaining variables are
irrelevent and can be discarded without much loss of predictive
or explanatory power. Doing this in a structured and principled way
usually requires algorithmic methods for choosing $X$. One such method,
forward stepwise regression, is a procedure
that begins with an empty model and sequentially adds the best predictor
variable in each step. Because of the stochastic nature of this
algorithm---making use of the data to choose variables---the
usual $\chi^2$ and $F$-tests for significance
fail when a model has been selected this way. These tests will be
anti-conservative unless they are computed on a held-out validation
dataset. This problem is one instance of the general problem of
conducting inference and model selection using the same data, a problem
of central importance on which some recent progress has been made.

In the LASSO setting, \cite{significance:lasso} derived
a novel test statistic and its asymptotic null distribution, making
possible valid inferences after model selection using the full data.
\cite{tests:adaptive} modified and extended those results to the
group LASSO \citep{grouplasso} and other penalized regression
problems, but only under the global null hypothesis.
One of the strengths of these test statistics is that they can be
used for valid significance testing when computed on the same
data used for model selection, eliminating the need for data splitting.
This is especially important in
situations in which data splitting is not appropriate. For example,
when categorical covariates have levels with very few observations
it can be difficult or impossible to split the sample with an adequate
number of such observations occurring in each split. Furthermore, even
if sample splitting is possible, it sacrifices accuracy in the model
selection procedure and power in the subsequent inferences. 

The present work iteratively applies the global null test
of \cite{tests:adaptive} for each step in forward stepwise selection,
and works out some of the details necessary for
models with grouped variables. The resulting method can
be more statistically efficient than validation on held-out data, and
more computationally efficient than penalized methods with
regularization parameters chosen by cross-validation.

As an illustrative example of what we gain from this method,
consider a response $Y$ with ten
categorical predictors $X_g$ having between 2 and 4 levels. The true 
relationship is that $Y$ only depends on $X_1$ and $X_9$. Specifically,
if observation $i$ has covariates $X_{1,i} = j, X_{9,i} = k$ then
\[
    Y_i = \beta_{1,j} + \beta_{9,k} + \epsilon_i , \qquad
    \epsilon_i \sim N(0,1)
\]
$X_1$ has three levels with $\beta_1 = (1, .5, -1)$, $X_9$
has two levels with $\beta_9 = (.5, -.5)$. We generated a random
categorical design matrix $X$ and created an instance of this example
with $n = 40$ observations. Running forward stepwise for eight steps,
we calculated the new $T\chi$ $p$-value, the usual
$\chi^2$ $p$-value based on drop in RSS at each step,
and an estimate of an exact $p$-value based on comparing the norm
achieved by the variable being added to the model, $\| X_{g^*}^T y \|_2$,
to quantiles of the Monte Carlo sample
\[
\max_{g \in A^c} \| X_g^T z_m \|_2, \quad z_m \sim N(0, I) \quad \text{for } m = 1,\ldots,M
\]
with $A^c$ being the set of variables not yet included
(updated \textit{after} computing this estimate, so that $g^* \in A^c$).

\begin{table}[ht]
\centering
\begin{tabular}{l|rrrrrrrr}
  \hline
Step & 1 & 2 & 3 & 4 & 5 & 6 & 7 & 8 \\ 
  \hline
Variable & 1 & 9 & 2 & 8 & 4 & 7 & 3 & 10 \\ 
  \hline
  $T\chi$ & 0.00 & 0.01 & 0.16 & 0.60 & 0.72 & 0.81 & 0.84 & 0.92 \\ 
  $\chi^2$ & 0.00 & 0.00 & 0.04 & 0.21 & 0.29 & 0.56 & 0.73 & 0.82 \\ 
  max-$\chi$ & 0.00 & 0.04 & 0.50 & 0.98 & 0.98 & 0.99 & 0.99 & 0.98 \\ 
   \hline
\end{tabular}
\caption{\em Small comparison of $p$-values. Elapsed time for forward stepwise \textit{and} $T\chi$ computation: 0.022 seconds, and for Monte Carlo sample estimate of max-$\chi$ with $M=200$: 0.235 seconds.} 
\label{tab:example}
\end{table}

In Table~\ref{tab:example} we see that forward stepwise chooses the
truly nonzero variables first and both $p$-values for these are small.
However, once forward stepwise begins adding noise variables, the
$\chi^2$ $p$-value remains small, potentially leading to 
incorrect inferences. The $T\chi$ and MC-estimated exact $p$-values
do not suffer from this selection effect.
However, the MC-estimate takes substantially more computational time to evaluate than the $T\chi$ statistic.

In the next section we establish general notation and describe the
forward stepwise algorithm used throughout the paper.
Section~\ref{sec:testing} reviews some recent work on
post-selection inference
\citep{significance:lasso,tests:adaptive,lasso:fixed}
relevant to our work here, and describes a general framework
for post-selection inference based on quadratic comparisons.
Simulation results in Section~\ref{sec:simulations} show
empirically that our method performs well in settings where forward
stepwise itself performs well, and that
various stopping rules using the $T\chi$ test statistic---including
some from \cite{sequential:fdr}---appear promising. In
Section~\ref{sec:applications} we apply
the method to several variants of forward
stepwise tailored to models with interactions and generalized additive
models, as well as to a real data example involving genomic prediction
of individual drug responses and resistances for various mutations of HIV.


\section{Forward stepwise model selection}
\label{sec:stepwise}

\subsection{Background and notation}

\begin{figure}
\begin{center}
\includegraphics[width=0.9\textwidth]{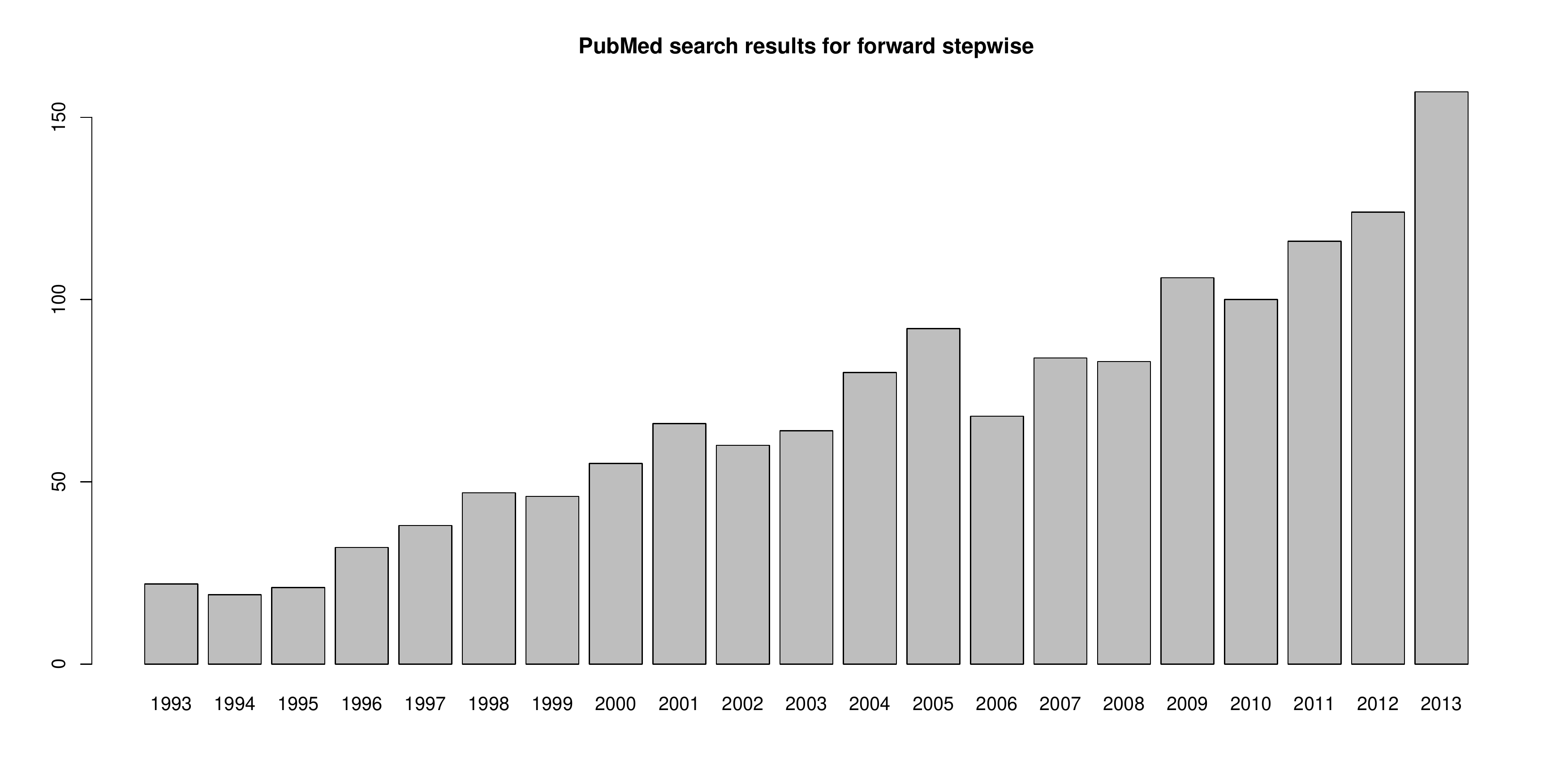}
\caption{\small \em PubMed search results show continued widespread usage of forward stepwise}
\label{fig:pubmed}
\end{center}
\end{figure}

As a classical method dating back about half a century
(see \cite{classical:selection} for a review),
forward stepwise regression has not received much attention in
recent years in the theoretical statistics community. But it continues
to be widely used by practitioners.
For example, search results on PubMed for forward stepwise,
summarized in Figure~\ref{fig:pubmed}, show that many recent
papers mention the method and there is an increasing trend over time.
Its popularity among researchers continues despite the fact that it
invalidates inferences using the standard $\chi^2$ or $F$-tests.

Some attempts to address this issue include Monte Carlo
estimation of tables of adjusted $F$-statistic values \citep{mc:ftoenter},
and permutation statistics \citep{permutation:stop}. Aside from the works
this paper is based on, there have been other recent attempts to do
inference after model selection. Most of these make use of subsampling
\citep{meinshausen:buhlmann} or data splitting \citep{wasserman:roeder}.
Our approach allows use of the full data and does not require the
extra computation involved in subsampling.
Before describing the full approach we first introduce notation and
specify our implementation of forward stepwise, which is slightly
different from the most commonly used versions.

We allow forward stepwise selection to add groups of variables in each
step, not only in the case of dummy variable encoding for categorical
variables but also for any grouping purpose. For example, groups of
variables may be pre-designated factors such as expression
measurements for all genes in a single functional pathway.
To emphasize this we use $g, h$ as covariate indices rather than the usual
$i, j$ throughout. Since single variables can be considered groups of
size 1, this includes non-grouped situations as a special
case.

Let $y \in \real^n$ be $n$ i.i.d. measurements of the outcome variable.
Let an integer $G \geq 2$ be the number of groups of explanatory variables.
For each $1 \leq g \leq G$ the design matrix encoding the
$g$th group is the $n \times p_g$ matrix denoted $X_g$, where $p_g$ is
the number of individual variables or columns in group $g$.
When a group encodes a categorical variable as indicators for the
levels of that variable, by default we use the full encoding with a
column for every level. Although this introduces collinearity, our method
does not require the design matrix to have full rank.

Denote by $p = \sum_{g=1}^Gp_g$
the total number of individual variables, so $p = G$ in the
case where all groups have size 1. Let $X$ be the matrix constructed
by column-binding the $X_g$, that is  
\begin{equation*}
X = \begin{pmatrix} X_1 & X_2 & \cdots & X_G  \end{pmatrix}
\end{equation*}
We also allow weights $w_g$ for each group. These
weights act like penalties or costs, so increasing $w_g$ makes it
more difficult for the group $X_g$ to enter the model. The
modeler can choose weights arbitrarily for calibration purposes,
but throughout we set them all constant (equal to 1) and normalize
groups by the Frobenius norm of their corresponding submatrices.

With each group we associate the $p_g \times 1$ coefficient vector
$\beta_g$, and write $\beta$ for the $p \times 1$ vector constructed
by stacking all of the $\beta_g$ in order.  Finally, our model for the
response is
\begin{equation}
\begin{aligned}
\label{eq:gmodel}
y & = X \beta + \sigma \epsilon \\
   & = \sum_{g=1}^G X_g \beta_g + \sigma \epsilon
\end{aligned}
\end{equation}
where $\epsilon$ is noise. We assume Gaussian noise
$\epsilon | X \sim N(0, \Sigma)$ with known covariance matrix $\Sigma$.

The model \eqref{eq:gmodel} is underdetermined when $p > n$.
In such cases it still may be possible to
estimate $\beta$ well if it is sparse--that is, if it has few nonzero
entries. In the rest of this paper we refer to variable groups $X_g$
as noise groups if $\beta_g$ is a zero vector and as true or signal
groups if $\beta_g$ has any nonzero entries. We refer to the number
of such nonzero groups as the \textit{sparsity} of the model, and
denote this $k := \# \{ g : \beta_g \neq 0 \}$. With this notation
we are now ready to describe our procedure concretely.

\subsection{Description of the forward stepwise algorithm}

First, the user must specify the maximum number of steps allowed,
which we denote \textit{steps}.
To enable $T\chi$ statistic computations,
\textit{steps} should be at most $\min (n, G) - 1$, but it is
computationally desirable to set it as low as possible while
safely larger than the sparsity of $\beta$. 
Then forward stepwise may recover all the nonzero
coefficients of $\beta$ and terminate without performing much
additional computation.
Of course the sparsity is usually unknown, so this requires guesswork.
In our implementation we treat the active set $A$ as an ordered list to easily track the order of groups entering the model.

\begin{algorithm}
  \caption{Forward stepwise variant with groups and weights}
  \label{algo:fs}
  \begin{algorithmic}[1]
    \REQUIRE An $n$ vector $y$ and $n \times p$ matrix $X$ of $G$ variable groups with weights $w_g$
    \ENSURE Ordered active set $A$ of variable groups included in the model at each step
    \STATE $A \gets \emptyset$, $A^c \gets \{ 1, \ldots, G\}$, $r_0 \gets y$
    \FOR{$s=1$ to $steps$}
    \STATE $g^* \gets \argmax_{g \in A^c} \{ \norm{X_g^Tr_{s-1}}_2 / w_g \}$
    \STATE $P_{g^*} \gets I_{n\times n} - X_{g^*}X_{g^*}^\dagger$
    \STATE $A \gets A \cup \{ g^* \}$, $A^c \gets A^c \backslash \{ g^* \}$
    \FORALL{$h \in A^c$}
      \STATE $X_h \gets P_{g^*} X_h$
    \ENDFOR
    \STATE $r_s \gets P_{g^*} r_{s-1}$
    \ENDFOR
    \RETURN $A$
  \end{algorithmic}
\end{algorithm}

The active set $A$ contains variable groups chosen to be included in
the model. Fitting $\hat \beta$ can be done by tracking the individual
fits and projections, or by simply fitting a linear model on the submatrix
of $X$ corresponding to $A$. 
 Note that other
implementations of forward stepwise use different criteria for choosing
the next variable to add, such as the correlation with the residual.
Since we do not renormalize the columns after projecting the covariates
(lines 6 to 8 above), and since we have weights, we are generally not computing
correlations unless the design matrix is orthogonal and all weights are 1. We could
renormalize the columns, though we choose not to.
There are advantages and disadvantages to both criteria which we do not
discuss. Our choice was motivated by the group LASSO result in
\cite{tests:adaptive}, but we believe other criteria can be handled
with appropriate modifications.
We now use forward stepwise to refer specifically to
Algorithm~\ref{algo:fs} unless otherwise specified.

\subsection{Performance of forward stepwise}
\label{sec:fs:performance}

Among model selection procedures, forward stepwise is one which performs
variable selection: from a potentially large set of variables it chooses
a subset to include in the model. The most ambitious form of variable
selection is  ``best-subset'' selection, a procedure which picks the
best model among all $2^G$ subsets of the $G$ possible groups.
This exhaustive search is computationally infeasible when $G$ is
large, and when possible it still
runs the risk of over-fitting unless model complexity is
appropriately penalized (as in \eqref{eq:subsetregress} below).
Forward stepwise produces a much
smaller set of potential models, with cardinality at most
\textit{steps} (which, recall, is less than $\min(n,G)$). However
it is a greedy algorithm, so the set of models it produces may not
contain the best possible model. This is an inherent shortcoming of
forward stepwise procedures and should be kept in mind when choosing
between model selection methods.

So far we have left open the question of choosing among the models in
the forward stepwise sequence, i.e. when to stop stepping
forward. Some approaches for this problem can be posed as optimization
criteria which stop at the step minimizing
\begin{equation}
\begin{aligned}
\label{eq:subsetregress}
\frac{1}{2} \| y - X \beta_s \|_2^2 + \lambda \pen(\beta_s)
\end{aligned}
\end{equation}
where we have written $\{ \beta_s : s = 1, \ldots, steps \}$ as
the sequence of models output by forward stepwise. The function
$\pen(\beta)$ is a penalty on model complexity usually taken to be the
number of nonzero entries of $\beta$. Proposals for $\lambda$ include
2 ($C_p$ of \cite{CP}, AIC of \cite{AIC}), $\log(n)$ (BIC of \cite{BIC}), and
$2\log(p)$ (RIC of \cite{RIC}). Stopping rules based on classical test
statistics have also been used, so it is natural to consider using
the new test statistics of \cite{significance:lasso} or
\cite{tests:adaptive}
to choose a model. \cite{sequential:fdr} examined some stopping rules
using the asymptotic $p$-values of \cite{significance:lasso} and showed
their stopping rules control false discovery rate--the expected
proportion of noise variables among variables declared significant
\citep{fdr}. We explore this further in Section~\ref{sec:simulations}.

Although forward stepwise is a greedy algorithm producing a
potentially sub-optimal sequence of models, under favorable conditions
it can still perform well. There is a subset of the compressed sensing literature \citep{donoho:pursuit, cai:wang:omp} dedicated to forward stepwise (often referred to in that literature as Orthogonal Matching Pursuit or OMP). 
Typical results from these works establish that forward stepwise can exactly select the true model under some stringent conditions involving quantities like the sparsity of the true model and the coherence of the design matrix.
The coherence $\mu(X)$ of a matrix $X$ with columns scaled to have unit 2-norm is defined as
\begin{equation}
  \mu := \mu(X) = \max_{i \neq j} \{ | \innerp{ X_i }{ X_j } | \}
\end{equation}
Denoting $k$ as the sparsity of $\beta$, the literature establishes
that if $k < (1/\mu + 1)/2$ and the nonzero coefficients of $\beta$ are sufficiently large then forward stepwise recovers $\beta$ with high probability.
The coherence condition is necessary to guarantee exact recovery \citep{cai:wang:xu:sharp} in the sense that it is possible to construct counterexamples with $k = (1/\mu + 1)/2$.
We refer the reader to \cite{donoho:pursuit, cai:wang:omp} for details.
For our purposes the conditions required to guarantee exact recovery are usually too stringent.
Simulations show empirically that forward stepwise can work well even when it is not working perfectly, and that it does so under a wide range of conditions.

\begin{figure}
\begin{center}
\subfigure[Categorical designs]{
\label{fig:fscat}
\includegraphics[width=0.6\textwidth]{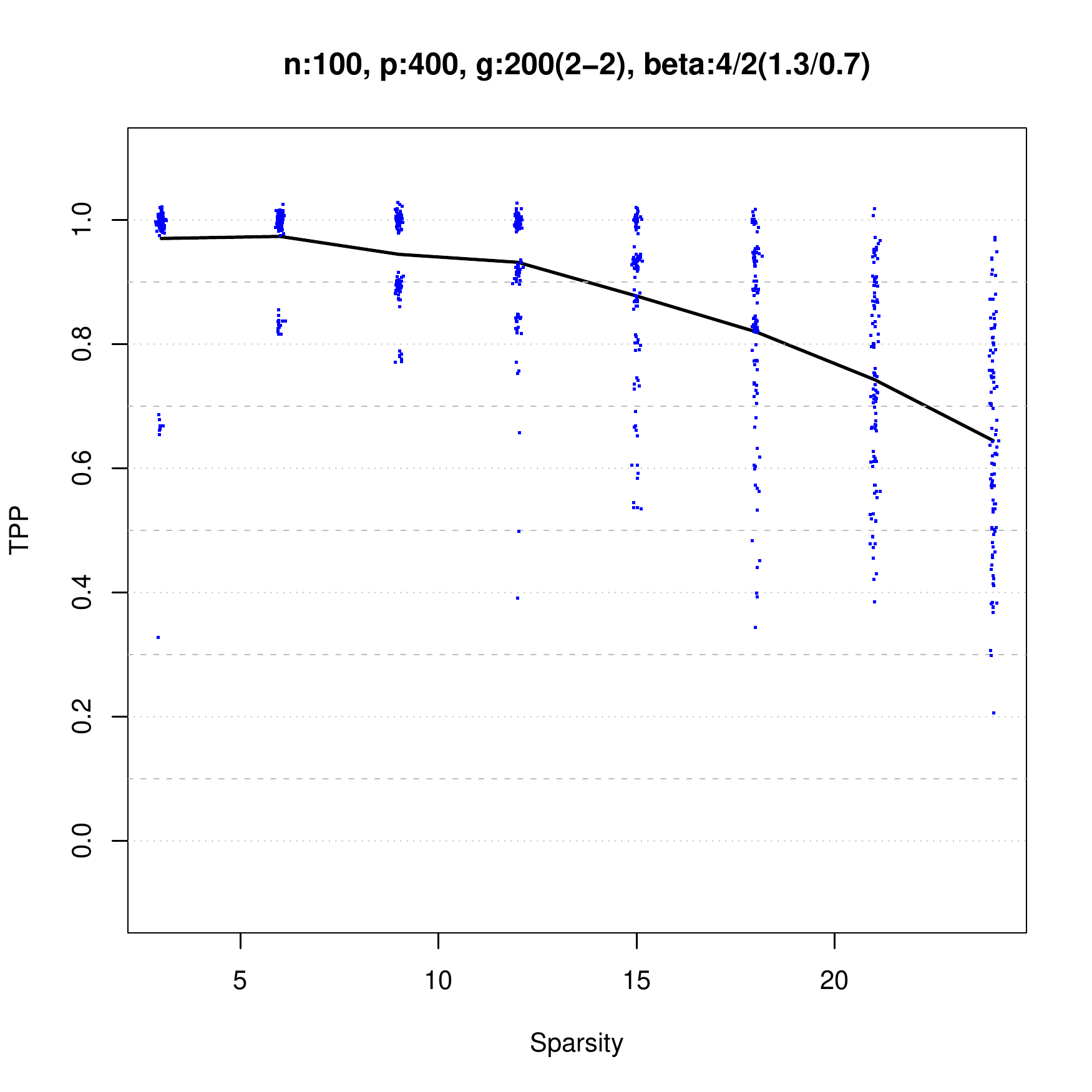}}
\hspace{-15pt}
\subfigure[Gaussian designs]{
\label{fig:fsgauss}
\includegraphics[width=0.6\textwidth]{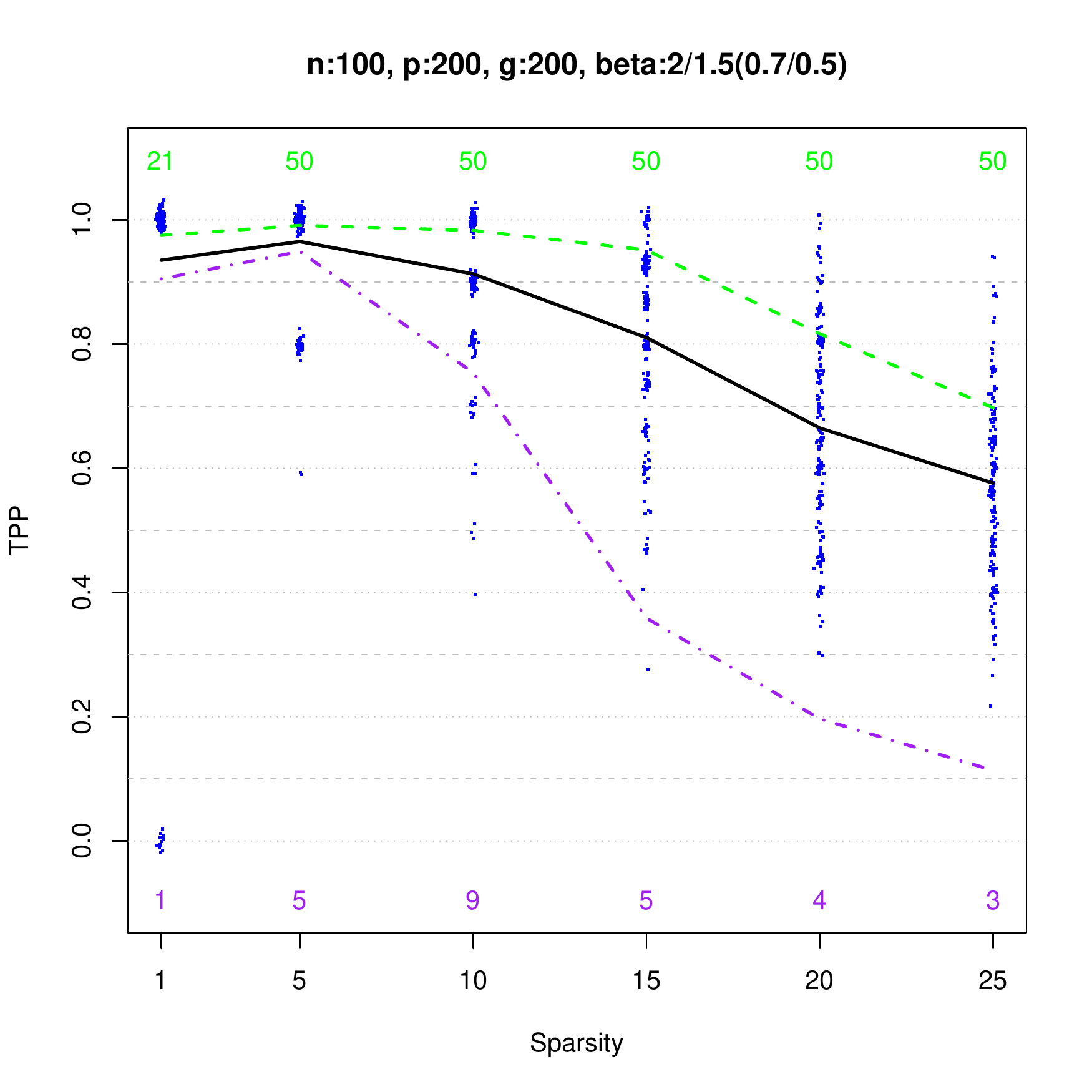}}
\caption{\small \em
  For various sparsity levels $k$, the True Positive Proportion of the
  model selected by forward stepwise at $k$ steps.
  The left panel shows the results of the simulation
  using a categorical design consisting of 200 binary factors.
  The right panel shows the results with design matrices of Gaussian
  entries. Plotted in green are the TPP (dashed line) and median model
  sizes (top) for models chosen by BIC, and in purple the same are shown
  (dot-dash line, bottom) for RIC.}
\label{fig:fwdstepsim}
\end{center}
\end{figure}

For various sparsity levels $k$, we applied Algorithm~\ref{algo:fs} to
various data sets with \textit{steps} set to $k$.
In the resulting active set $A_k$, the number of true variables divided by $k$ is the $k$-True Positive Proportion ($k$-TPP):
\[
k\text{-TPP} = \frac{\# \left\{g: g \in A_k, \beta_g \neq 0\right\}}{k}.
\]
Since we know $k$ in the simulations we can use $k$-TPP as a measure of how well forward stepwise is performing. When it is close to 1 we are recovering most of the true variables before including false ones.
In our simulations nonzero coefficients have magnitudes
in a range of multiples of $\gamma := \sqrt{2\log(G)/n}$, e.g. in $[1.4
\gamma, 1.6 \gamma]$. Results are shown in
Figure~\ref{fig:fwdstepsim}. After computing the
coherence of these design matrices, some calculations show the required
sparsity level to guarantee exact recovery in these situations is
about 2 or smaller, and the required nonzero coefficient magnitude is
likely in the range of 10 to 100 times $\gamma$. These simulations are far
from the stringent conditions required by theory to guarantee perfect
recovery, but the performance, while not perfect, may still be good
enough for some purposes. Finally, to serve as a comparison with existing
commonly used model selection procedures, in Figure~\ref{fig:fsgauss}
we show in green the TPP and median model sizes for models chosen by
the \textit{step} function in \textit{R} using the BIC criterion, and
the same for RIC are plotted in purple.


\section{Significance testing with model selection}
\label{sec:testing}

\subsection{Background}
In the ordinary least squares setting, a significance test for a
single variable can be conducted by comparing the drop in residual
sums of squares (RSS) to a $\chi^2_1$ distribution. Similarly, when
adding a group of $k$ variables we can compare the drop in RSS to a
$\chi^2_k$ random variable. This generally does not work when the
group to be added has been chosen by a method that uses the data
\citep{olshen:flevel},
and in particular it fails for forward stepwise procedures adding
the ``best'' (e.g. most highly correlated) predictor in each step. In
that case, the drop in RSS does not match its
theoretical null distribution even when the null hypothesis is
true. \cite{significance:lasso} introduced a new test statistic based
on the knots in the LASSO solution path. They derived a simple
asymptotic null distribution, proved a type of convergence under broad
``minimum growth'' conditions, and demonstrated in simulations that
the test statistic closely matches its asymptotic distribution even in
finite samples.  That work marked an important advance in the problem
of combining inference with model selection.  \cite{tests:adaptive}
extended that work to the group LASSO \citep{grouplasso} and other
problems, and modified the test statistic to one with an exact finite
sample distribution under the global null hypothesis.

To describe the previous work we require some facts about the solution path
of the LASSO.
Recall the LASSO estimator is given by
\begin{equation}
\begin{aligned}
\label{eq:lasso}
\displaystyle \hat \beta(\lambda) = \argmin_{\beta \in \real^p} \frac{1}{2} \| y - X \beta \|_2^2 +
   \lambda \| \beta \|_1 \\
\end{aligned}
\end{equation}
The following facts are summarized in \cite{significance:lasso,tibshirani_LASSO_uniqueness}.

\begin{itemize}

  \item The vector valued function $\hat \beta(\lambda)$ is a
    continuous function of $\lambda$. For the LASSO path, the
    coordinates of $\hat \beta(\lambda)$ are piecewise linear with
    changes in slope occurring at a finite number of $\lambda$ values
    referred to as \emph{knots}. 
    \item The knots depend on the data and are
    usually written in order $\lambda_1 \geq \lambda_2 \geq \cdots
    \geq \lambda_r \geq 0$. 

\end{itemize}

Assuming groups of size one and $\Sigma = \sigma^2 I$, the covariance test is given at the first step by
\begin{equation}
\label{eq:covtest}
T_1 = \frac{\lambda_1(\lambda_1-\lambda_2)}{\sigma^2} \overset{n,p \to \infty}{\to} \text{Exp}(1).
\end{equation}
This is a hypothesis test for including the first variable in the LASSO
path, with large values of the test statistic being evidence against the
global null.
In \cite{tests:adaptive} it is pointed out 
\begin{equation}
\label{test:exact}
\exp\left(- \frac{\lambda_1(\lambda_1-\lambda_2)}{\sigma^2} \right) \approx \frac{1 - \Phi(\lambda_1/\sigma)}{1 - \Phi(\lambda_2 / \sigma)} \overset{D}{=} \text{Unif}([0,1]).
\end{equation}
The right hand side has an exact, finite sample null distribution.
Further, this new statistic can be understood as the survival function of $\lambda_1 = \|X^Ty\|_{\infty}$
{\em conditional} on which variable achieves $\lambda_1$ as well as its sign.
The limiting results about later steps in \cite{significance:lasso} can
be interpreted as
recursively applying $T_1$ to the variables that had not previously been selected by LARS. In forward stepwise with groups of size 1, we recursively apply the right hand side test statistic in the same manner. With larger groups, we use an analogous test statistic to the right hand side of \eqref{test:exact}. The test statistic is presented as an example in \cite{tests:adaptive} though we re-derive it here in simpler form. The test draws inspiration
from the group LASSO.

The \emph{group LASSO estimator} \citep{grouplasso,bakin:grouplasso},
a generalization of the LASSO, is a solution to the following penalized
least-squares convex problem
\begin{equation}
\begin{aligned}
\label{eq:gsoln}
\displaystyle \hat \beta_\lambda = \argmin_{\beta \in \real^p} \frac{1}{2} \| y - X \beta \|_2^2 +
   \lambda {\cal P}(\beta) \\
\end{aligned}
\end{equation}
with the group penalty
\begin{equation}
  \begin{aligned}
  \label{eq:gpen}
    {\cal P}(\beta)&= \sum_{g=1}^G w_g \| \beta_g \|_2 \\
  \end{aligned}
\end{equation}
The parameter $\lambda \geq 0$ enforces sparsity in groups: for large
$\lambda$ most of the $\beta_g$ will be zero vectors. The weights
$w_g$ are usually taken to be $\sqrt {p_g}$ to normalize the penalty
across groups with different sizes.  This can also be accomplished
by scaling the corresponding submatrices by their Frobenius norms
and setting all weights equal.
Note that this includes the usual LASSO estimator as a
special case when all groups are of size 1, since then the
penalty term is the $\ell_1$-norm of $\beta$. The solution path of the group
LASSO has similar properties to the LASSO case, however it is not
generally piecewise linear.

For sufficiently large $\lambda$ the solution to~\eqref{eq:gsoln} is forced
to be zero. The smallest such $\lambda$ is denoted
\begin{equation}
  \begin{aligned}
   \lambda_1 &= \inf \{ \lambda \geq 0 : \hat \beta_{\lambda'} = 0 \text{ for all } \lambda' > \lambda \} \\
  \end{aligned}
\end{equation}
For the group LASSO this is explicitly computable by
\begin{equation}
\lambda_1^{\text{group}} = \max_{g} \frac{1}{w_g}\|X_g^Ty\|_2 . 
\end{equation}
The value above is also the dual norm of the penalty~\eqref{eq:gpen}
hence it can be expressed as
\begin{equation}
\label{eq:lammax}
\lambda_1^{\text{group}}
 = \frac{1}{w_{g^*}}\eta_{g^*}^TX_{g^*}^Ty . 
\end{equation}
where $g^*$ is the group that achieves the maximum and 
$\eta_{g^*} = X_{g^*}^Ty / \|X_{g^*}^Ty\|_2$ is the unit vector that achieves the norm $\|X_{g^*}^Ty\|_2$. One of the key conceptual points about our
test is that it conditions on both the maximizer $g^*$ and on the unit
vector $\eta_{g^*}$. Most of the $p$-value computations will be done in
terms of these quantities.

\subsection{Quadratic framework and derivation of test statistic}

\begin{figure}[!htp]
\begin{center}
\resizebox{!}{3in}{\input{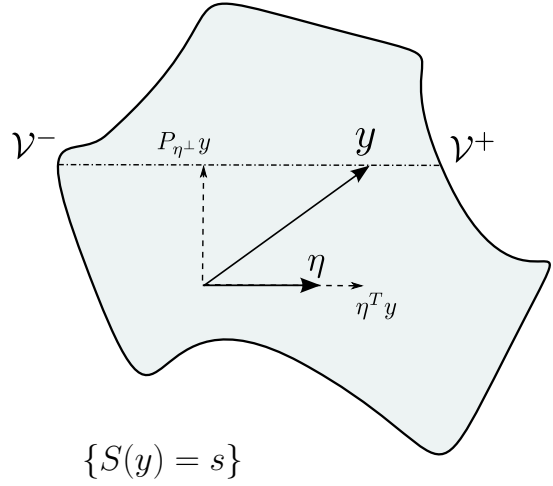}}
\end{center}
\caption{ \em
A quadratic selection event, given by the intersection of a list
of quadratic inequalities. The selection event and chosen contrast
$\eta$ determine a slice through the event with computable limits
where the boundary of the event is reached.
Note that the event need not be simply connected.
}
\label{fig:curved}
\end{figure}

In this section we derive the new $p$-value and describe how to compute it.
First we give a brief summary of what follows; readers not interested in the full derivation can skip the rest of this section after they understand the summary. The event $E_g$ that forward stepwise Algorithm~\ref{algo:fs} selects a given group $g$ is equivalent to a set of quadratic inequalities involving $y$ (see \eqref{eq:first:quadratic} below). To derive a statistic related to the group $g$ being included, we choose a direction vector $\eta(g)$ in the relevant direction and study the distribution of $\eta(g)^Ty$ restricted to the event $E_g$. In Figure~\ref{fig:curved} below the event $E_g$ is the shaded region, and the direction $\eta$ determines a slice through this event. Finally, we solve for the amounts $t$ by which $y$ can be translated in the directions $\pm \eta$ and still satisfy the constraints imposed by $E_g$. These are the points where the slice intersects the boundary of $E_g$. Let $M \subset \mathbb R$ be the set of $t$ such that $y + t\cdot\eta \in E_g$. With all of these quantities and a little more work we find that the observed norm $\| X_g^Ty \|_2$ has a $\chi$ distribution truncated to the set $M$, and with a computable scale parameter. Applying the appropriate CDF transform \eqref{eq:tchi} with these quantities yields our $p$-value.

We now give the full derivation of the $p$-value, beginning with a previous work and then extending the framework there to the setting with groups.
The approach to post-selection inference in \cite{lasso:fixed}
describes the selection event for the LASSO as a convex polytope
satisfying a list of affine constraints.
If all groups are of size 1, then
the event that we observe a given set of variables
in the forward stepwise path (with their signs as they enter) would similarly
be given by a set of affine inequalities.
After selection, exact inference for a (randomly) chosen contrast $\eta(y)^T\mu$ of the mean vector
$\mu$ could then be accomplished by analyzing the conditional distribution
\begin{equation}
y \sim N(\mu, \sigma^2 I) \bigl| Ay \leq b.
\end{equation}
See \cite{lasso:fixed,exact:lars} for further details on this approach.

However, with group sizes larger than 1, the event that the first group
chosen $g^*(y)$ is equal to some fixed group $g$ is no longer given by
a set of affine constraints. Rather, considering line 3 of
Algorithm~\ref{algo:fs}, we see
\begin{equation}
\label{eq:first:quadratic}
\begin{aligned}
\{g^*(y)=g\} &= \{ \|X_g^Ty\|_2 /w_g \geq \|X_h^Ty\|_2 /w_h , \ \forall h \neq g\} \\
&=  \{ y^T(X_g^TX_g)y / w_g^2 - y^T(X_h^TX_h)y / w_h^2 \geq 0, \ \forall h \neq g\} \\
\end{aligned}
\end{equation}
Hence, our selection event can be expressed as 
the intersection of a list of quadratic inequalities. This non-affine selection event is stylized in Figure
\ref{fig:curved}.
In the rest of this section, we consider an arbitrary selection procedure
$S$ that returns one of a set of possible outcomes $s \in {\cal S}$ determined by a set of quadratic inequalities.
That is, for all possible outcomes $s \in {\cal S}$ there is a set of indices
$I(s)$ such that
\begin{equation}
\label{eq:selection}
\begin{aligned}
\{S(y)=s\} = \cap_{i \in I(s)} \{y: y^TQ_iy + a_i^Ty\leq b_i \}.
\end{aligned}
\end{equation}
The quadratic forms are not assumed non-negative definite, but we can, without loss of generality
assume they are symmetric.
As in \cite{lasso:fixed} we will be interested in some randomly chosen contrast
$\eta(s)^T\mu$.
In our grouped selection procedure, the selection rule is $S(y)=g^*(y)$
and the value of $\eta$ on $\{ g^* = g \}$ is
$$
\eta(s) = \eta(g) = \frac{X_{g}^Ty}{\|X_{g}^Ty\|_2}.
$$
This choice of $\eta$ reflects our goal of calculating a $p$-value for inclusion of the variable $g^*$ chosen by forward stepwise.
Note that for each fixed $g$ and any $\eta \in \real^g$ with $\|\eta\|_2=1$, the event 
$$
\left\{y:\frac{X_g^Ty}{\|X_g^Ty\|_2} = \eta \right\} = \left\{y: \|X_g^Ty\|^2_2 \leq (X_g\eta)^Ty \right\},
$$
is defined by another quadratic inequality that can be appended to $I(s)$.

Having fixed $\eta$, we proceed to study this contrast by slicing through the selection event along a ray with direction $\eta$ that passes through $y$. 
That is, we need to find
$$
\begin{aligned}
\lefteqn{
\left\{t \in \mathbb R : S(y+t \cdot \eta)=s \right\} } \\
 & \qquad = \cap_{i \in I(s)} \left\{t \in \mathbb R : (y+t \cdot \eta)^T Q_i (y+t \cdot \eta) + a_i^T(y+t \cdot \eta) \leq b_i\right\}. \\
\end{aligned}
$$
The slice for any particular inequality is
\begin{equation}
\begin{aligned}
\label{eq:quadslice}
\lefteqn{
\left\{t \in \mathbb R : (y+t\cdot \eta)^T Q_i (y+t \cdot \eta) + a_i^T(y+t \cdot \eta) \leq b_i\right\}} \\
 & \qquad = \left\{t \in \mathbb R : t^2 \cdot (\eta^TQ_i\eta) + t \cdot(2 y^TQ_i\eta + a_i^T\eta) + y^TQ_iy + a_i^Ty - b_i \leq 0 \right\} \\
\end{aligned}
\end{equation}
This can be computed explicitly, and results in either the empty set, a single interval (possibly infinite) or the union of 
two disjoint infinite intervals. The intersection over all $I(s)$ is therefore also computable, yielding a form for the slice
\begin{equation}
\label{eq:slice}
\{g^*(y)=g, \eta(g^*(y))=\eta\}.
\end{equation}
In Figure \ref{fig:curved}, for every $\eta$, the slice \eqref{eq:slice} is a function
of $P_{\eta}^{\perp}y$. This amounts to a proof of the following lemma.

\begin{lemma}
Suppose $S$ is a selection procedure, i.e. a map $S:\real^n \mapsto {\cal S}$
such that for each $s \in S$ \eqref{eq:selection} holds 
and we are given a matrix valued function $X_s \in \real^{n \times p(s)}$ of rank
$k(s)$.
Then, for every $\eta_s \in \real^{p(s)}$ with $\|\eta_s\|_2=1$, the
slice
\begin{equation}
\label{eq:slice:general}
\left\{t: S(y+t \cdot \eta_s)=s, \frac{X_s^Ty}{\|X_s^Ty\|_2}=\eta_s \right\}
\end{equation}
can be described by
a finite union of closed intervals whose endpoints are functions of 
$(P_s^{\perp}y, (P_s- (X_s\eta_s)(X_s\eta_s)^T)y)$ where
$P_s^{\perp}$ is projection on the orthogonal column space of $X_s$.
\end{lemma}
Denote the set \eqref{eq:slice:general} by $E(s, \eta_s, P_s^{\perp}y, (P_s - X_s \eta_s(X_s\eta_s)^T)y)$ and also define
$$
\begin{aligned}
\lefteqn{
E_+(s, \eta_s, P_s^{\perp}y, (P_s - X_s \eta_s(X_s\eta_s)^T)y) } \\
 & \qquad = \left\{y + t \cdot \eta_s: S(y+t \cdot \eta)=s, \frac{X_s^Ty}{\|X_s^Ty\|_2}=\eta_s \right\} \\
\end{aligned}
$$

We now present an explicit algorithm for computing this slice in quadratic selection procedures.

\begin{algorithm}
 \caption{Truncation interval for quadratic decisions}
 \label{algo:quadratic}
 \begin{algorithmic}
   \REQUIRE Response $y$, state $s$ list of quadratic inequalities $\{Q_i, a_i, b_i: i \in I(s)\}$, direction
   of interest $\eta$ with $\|\eta\|_2=1$.
   \ENSURE The set $\{t: (y+t \cdot \eta)^TQ_i(y+t \cdot \eta) + a_i^T(y+t \cdot \eta) \leq b_i  \ \forall i \in I(s) \}$.
      \STATE Initialize  interval: $M= (-\infty,\infty)$, $U=-\infty, L=\infty$
    \FOR{$i$ in $I(s)$}
    \STATE{$a=\eta^TQ_i\eta, b=2y^TQ_i\eta + a_i^T\eta, c=y^TQ_iy + a_i^Ty - b_i$}
    \IF{$a \neq 0$}
    \IF{$b^2-4ac > 0$}
    \STATE Stop: $y$ does not satisfy inequalities!
    \ELSIF{$a > 0$}
     \STATE $M \gets M \cap [(-b-\sqrt{b^2-4ac})/2a, (-b+\sqrt{b^2-4ac})/2a]$;
     \ELSIF{$a < 0$}
     \STATE $L \gets \min(L,(-b-\sqrt{b^2-4ac})/2a)$,
     \STATE $U \gets \max(U,(-b+\sqrt{b^2-4ac})/2a).$;
    \ENDIF
    \ELSE
    \IF{$b > 0$}
   \STATE $M \gets M \cap (-\infty, -c/b]$;
   \ELSIF{$b < 0$}
   \STATE $M \gets M \cap [-c/b,\infty)$;
   \ELSIF{$c > 0$}
   \STATE Stop: $y$ does not satisfy inequalities!
   \ENDIF
   \ENDIF
   \ENDFOR
   \RETURN $(M \cap (L,U)^c) + \eta^Ty.$
 \end{algorithmic}
\end{algorithm}

In turn, this allows us to derive a test statistic to test the hypothesis
$H_{0,s}:X_s^T\mu=0$ conditional on $s$ being selected.
\begin{lemma}
\label{eq:test:dbn}
For $y \sim N(0, \sigma^2 I)$,
conditional on the {\em event}
\[
H_{0,s} \cap E_+(s, \eta_s, P_s^{\perp}y, (P_s - \eta_s\eta_s^T)y)
\]
we have the following truncated $\chi$ distributional result
\begin{equation}
\label{eq:chi:truncated}
(X_s\eta_s)^Ty \overset{D}{=} \theta_s \chi_{k(s)} | (X_s\eta_s)^TE_+(s, \eta_s, P_s^{\perp}y, (P_s - \eta_s\eta_s^T)y) 
\end{equation}
where the vertical bar $|$ here denotes restriction to an interval, and the scale is given by
$$
\theta_s = \sigma \frac{y^TP_sy}{y^TX_sX_s^Ty}.
$$
\end{lemma}

{\bf Proof:}
For any fixed $s$, decompose $y$ as
$$
(Z^{\perp}_s,Z_s) = (P_s^{\perp}y, X_s^Ty).
$$
Under $H_{0,s}:X_s^T\mu=0$, the density of $Z_s$ can be written as
$$
(2 \pi \Sigma_s \sigma^2)^{-k(s)/2} \exp \left(-\frac{1}{2 \sigma^2} z^T\Sigma_s^{-1}z \right)
$$
with $\Sigma_s = X_s^TX_s$.
Transforming to polar coordinates $(R,U)$ yields a density
$$
(2 \pi \Sigma_s \sigma^2)^{-k(s)/2} r^{k(s)-1} \exp \left(-\frac{r^2}{2 \sigma^2} u^T\Sigma_s^{-1}u \right), \qquad r \geq 0, \|u\|_2=1.
$$
Conditioning on $U$ shows that $R=\|X_s^Ty\|_2|U$ has distribution proportional to a $\chi_{k(s)}$. Finally, observe that 
the scale of the $\chi$ above is given by $\theta_s$. \qed

\subsection{The $T\chi$ test statistic}

With Lemma \ref{eq:test:dbn} and Algorithm \ref{algo:quadratic} we are finally ready to define our test statistic, which we term the {\em truncated $\chi$ test statistic}, denoted $T\chi$. This statistic is determined by a degrees of freedom parameter $r$ as well as a truncation set $M \subset \real$, a scale parameter $\theta$ and an observed value $t$. The test statistic is given by the survival function of a $\chi_r$ random variable truncated to the set $M$ evaluated at our observed norm $\|X_g^Ty\|_2$ (after scaling by the appropriate scale $\theta$).
\begin{equation}
\label{eq:tchi}
T\chi(t, r, \theta,  M) = \frac{\int_{M /\theta \cap [t/\theta,\infty)} F_{\chi_r}(dz)}{\int_{M / \theta} F_{\chi_r}(dz)}.
\end{equation}
For simplicity we always use this distributional transform rather than reporting the observed values of $(t,r,\theta,M)$. Hence, $T\chi$ itself is our $p$-value, with
\[
T\chi \sim \text{Unif}([0,1])
\]
under the global null.

\begin{figure}[!htp]
\begin{center}
\subfigure[Gaussian with correlations]{
\label{fig:gausscorr}
\includegraphics[width=0.6\textwidth]{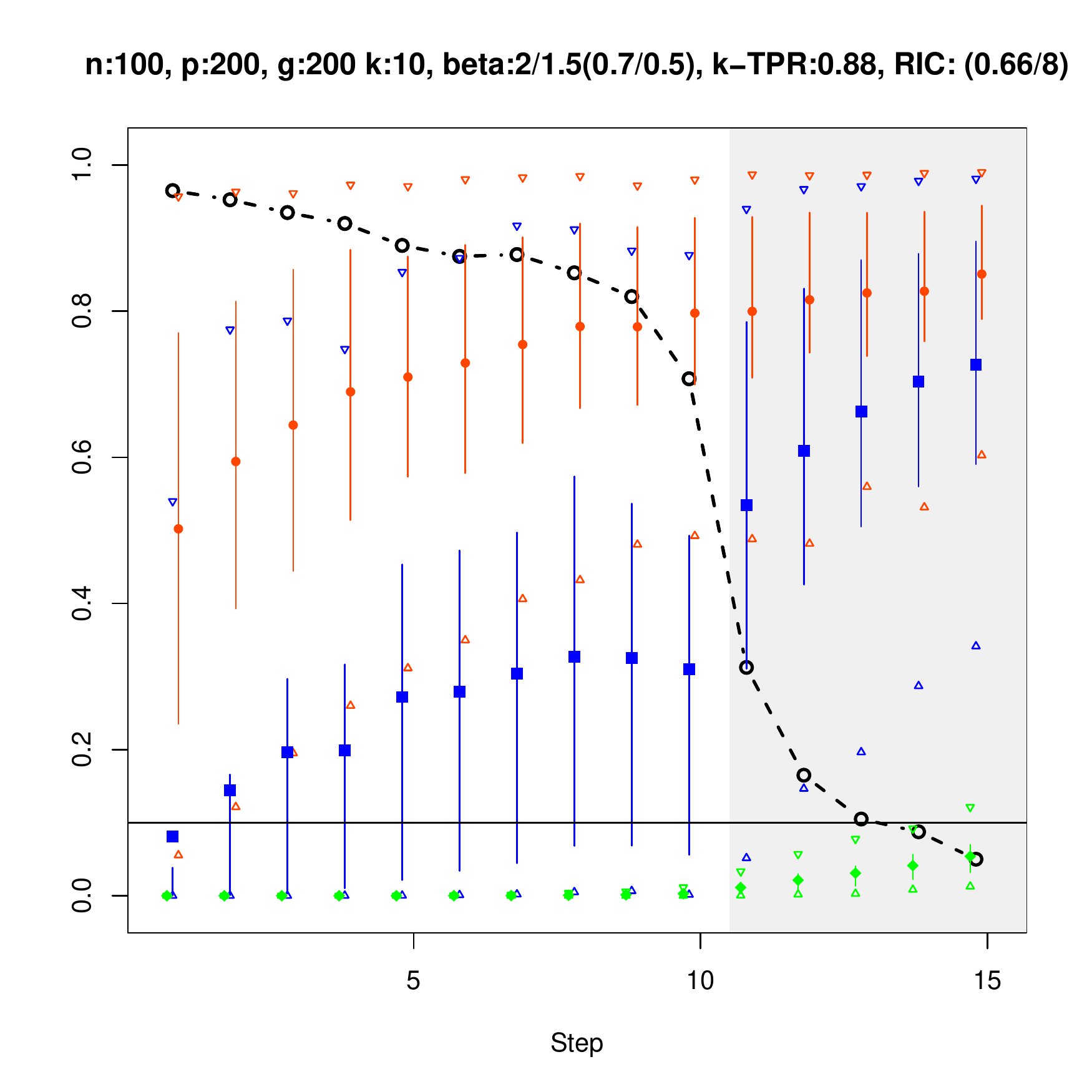}}
\hspace{-15pt}
\subfigure[Gaussian with groups]{
\label{fig:gaussgroups}
\includegraphics[width=0.6\textwidth]{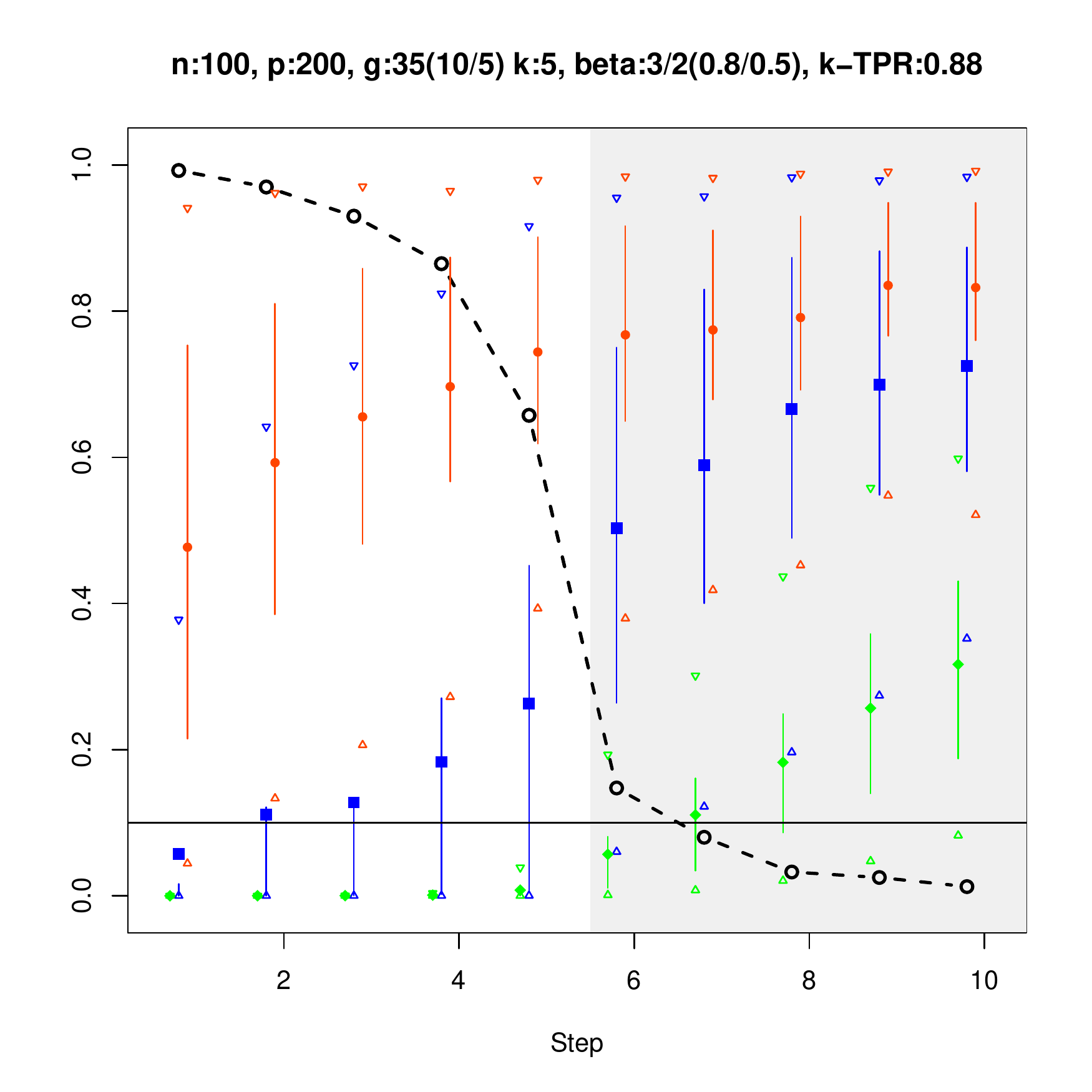}}
\caption{\small \em
Forward stepwise with $T\chi$ simulation results.
Dashed line is the local True Positive Rate at each step averaged over
400 simulation realizations.
Vertical lines
are boxplots of $p$-values marginally at each step, with red computed
using $T\chi$ on a null model, blue using $T\chi$ on a sparse alternative
model, and green using $\chi^2$ on the same. The shaded region indicates
the sparsity of the alternative.
Left panel shows results for Gaussian design
matrices with equicorrelation $\rho=0.2$ and equicorrelated noise with
correlation 0.1. Right panel shows results for an independent
Gaussian design with groups of sizes 5 and 10.}
\label{fig:pval1}
\end{center}
\end{figure}

After adding the first group in forward stepwise and computing the $T\chi$ $p$-value, we orthogonalize the response $y$ and all remaining groups with respect to the group just added. This imposes further constraints that in principle should be used when computing $p$-values at subsequent steps. For now we ignore these constraints and iterate the global null test, but work on incorporating all known constraints is ongoing. We believe that the effect of not tracking all constraints is what causes the nominal $p$-value to be increasingly stochastically larger than uniform further down the forward stepwise path (as can be seen in Figure~\ref{fig:pval1}), and only truly uniform at the first step where all remaining variables are noise.

For an alternate derivation that follows more closely \cite{tests:adaptive}, and an algorithm that includes weights and an arbitrary (known) covariance matrix $\Sigma$ see Appendix~\ref{app:algo}.

\section{Simulations}
\label{sec:simulations}

To study the behavior of the $T\chi$ test after taking steps in the forward stepwise path and to understand its power to detect various departures from the global null we conduct simulations in a wide variety of settings including both $p > n$ and $p < n$ problems.
We performed simulations with several classes of random design matrices
including Gaussian and categorical, and with fixed categorical designs
from the HIV data set of Section~\ref{sec:hiv}. Gaussian
design matrices were generated either independently or with some global
correlation $\rho > 0$ between all pairs of columns. Categorical matrices
were generated by first choosing a vector of probabilities for a given
variable from a Dirichlet prior, and then sampling category levels with
that vector of probabilities. Resampling was used to ensure the minimum
number of observations in any level was at least five. Finally, categorical
variables were encoded as groups of 0-1 vectors using the full encoding.

Since we are interested in variable selection we generate signals with
nonzero coefficients on the scale of $\gamma := \sqrt{2 \log(G)/n}$,
where recall $G$ is the number of groups.
Coefficients within a
nonzero group have roughly the same magnitude and are normalized
so that $\| \beta_g \|_2$ has scale $\gamma$ independently of $p_g$.
The magnitudes for different
nonzero groups range from a lower limit times $\gamma$ to an upper
limit times $\gamma$. In text above each plot, the limits are listed
next to ``beta'' and the numbers in parentheses are the largest
and smallest 2-norms of nonzero coefficient groups respectively.
Each plot also displays the
number of observations or rows, $n$, the number of columns, $p$, the
number of groups, $g$, and the largest and smallest group sizes in
parentheses if the groups are not all size 1. The number of nonzero
groups, $k$, is also displayed graphically by shading with the portion
of the plot showing steps after $k$ shaded gray.
The proportion of truly
nonzero groups recovered in the first $k$ steps is denoted $k$-TPP,
where recall $k$ is the number of truly nonzero groups. 

We show two types of plots, one containing the same information as in
Figure~\ref{fig:fwdstepsim}. The other plots each show a scenario with
a fixed sparsity level and contain the following information. The
horizontal axis is the step index for forward stepwise. The dashed
line shows the proportion of iterations where a truly nonzero variable
was added to the model at that step. Solid verticle lines are essentially
narrow boxplots, showing the middle 50\% of $p$-values calculated at that
step with circular points in these lines as the average and triangles
showing the 95\% and 5\% quantiles. Different color verticle lines
represent the following: red are calculated using the $T\chi$ $p$-value
on a null model with no nonzero groups, blue are calculated using the
$T\chi$ $p$-value on the non-null model, and green calculated on the
non-null model using the usual $\chi^2$ significance test.

\begin{figure}[!htp]
\begin{center}
\subfigure[Real data categorical design]{
\label{fig:hivnrti}
\includegraphics[width=0.6\textwidth]{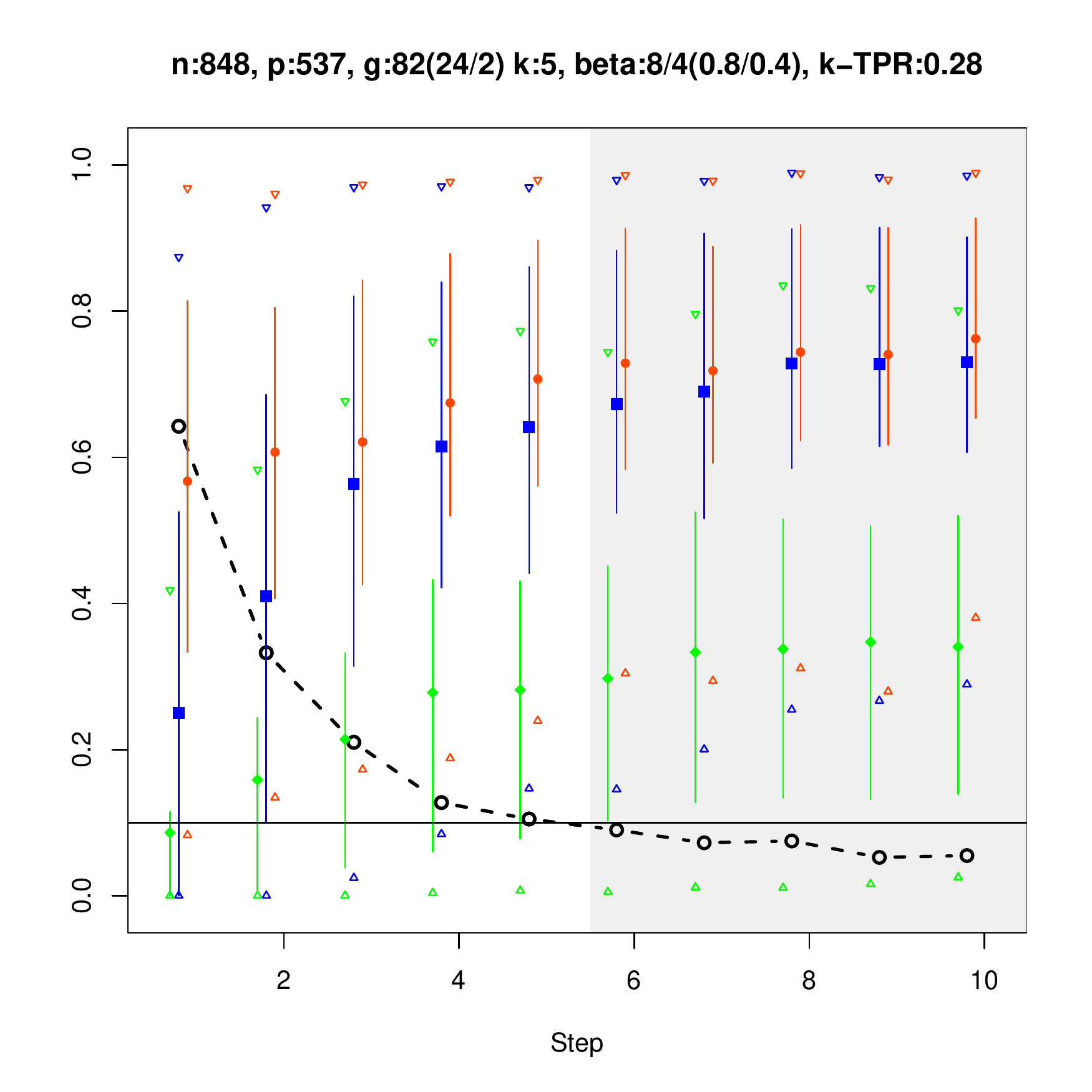}}
\hspace{-15pt}
\subfigure[Simulated categorical design]{
\label{fig:gausscat}
\includegraphics[width=0.6\textwidth]{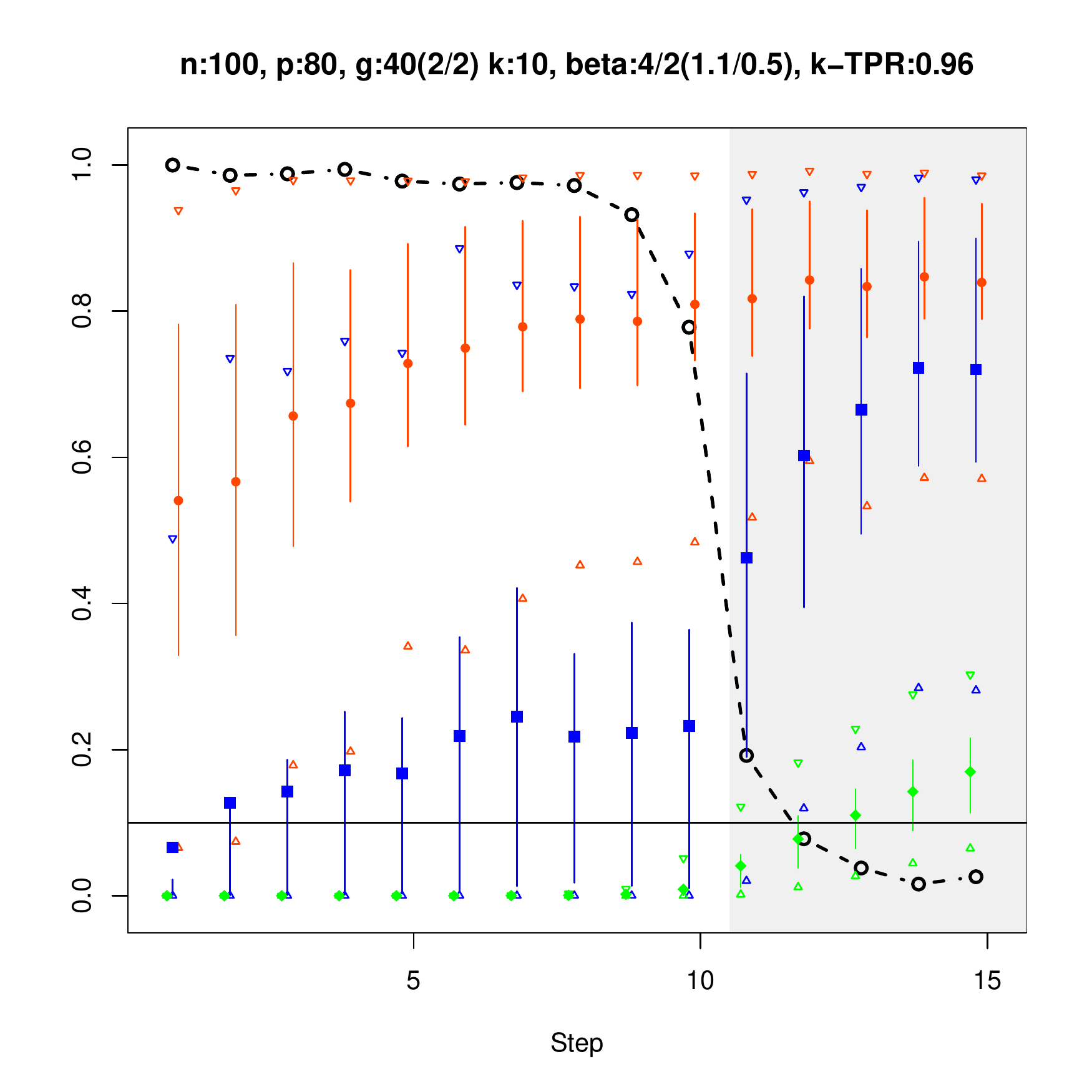}}
\caption{\small \em
As in Figure \ref{fig:pval1}.
The left panel shows results for a categorical design
taken from a real data set.
The right panel shows results for categorical matrices with all categories having 5 levels. The null cases being larger than uniform are likely the result of numerical error as these problems are poorly conditioned.
}
\end{center}
\end{figure}

The performance of forward stepwise in these plots is characterized by how
close the dashed line is to a step function. If forward stepwise finds all
truly nonzero variables first this line will be close to 1 until the step
index reaches $k$ and then it will be close to 0. When this is the case,
the $T\chi$ $p$-value tends to be small while forward stepwise is selecting truly
nonzero variables, uniform on the step where the dashed line goes to 0,
and then progressively larger afterward. A reasonable stopping rule based
on this behavior is to pick the model including all variables up to and
including the last one with a significant $p$-value. We call this the
\textit{last} stopping rule, and it selects the first $\hat k_\textit{last}$
variables in the forward stepwise path where, if
$p_j$ denotes the $T\chi$ $p$-value calculated at the $j$-th step, then
\begin{equation}
  \label{eq:klast}
  \begin{aligned}
    \hat k_\textit{last} & = \max \{ j \leq steps : p_j < \alpha \}. \\
  \end{aligned}
\end{equation}
To test this stopping rule we compare it with several others which we now
describe. The \textit{oracle} stopping rule assumes knowledge of the
sparsity $k$ and always picks the first $k$ variables. The \textit{first}
rule stops immediately before the first variable to yield a $p$-value
larger than $\alpha$, and since this requires rejection of the global 
null in order to include the first variable
it controls the family-wise error rate. However it does not seem to have
good power.
The \textit{forward}
stopping rule is defined in \cite{sequential:fdr} and has the desirable
property that it controls the false discovery rate. Finally, \textit{RIC}
and \textit{BIC}
are the risk inflation criterion and Bayesian information criterion
described in Section~\ref{sec:fs:performance}. Tables~\ref{tab:easy} and
\ref{tab:hard} show simulation results with $\alpha = 0.1$.
To summarize these results, \textit{forward} is the only rule which
controls FDR, \textit{BIC} selects models that are far too large,
\textit{RIC} has low FDR and good power when the sparsity is low,
and \textit{last} is comparable to \textit{RIC},
with perhaps more power in the higher dimensional setting
(and a correspondingly larger FDR when the signal is also weak).

\begin{table}[ht]
\centering
\begin{tabular}{|r|r|lll|lll|lll|}
 \hline
   \multicolumn{2}{|c|}{p} & \multicolumn{3}{c|}{50}  & \multicolumn{3}{c|}{500} \\ 
  \hline \hline
    k & rule  &  R        & FDP        &  TPP          &   R      & FDP        & TPP \\ 
  \hline 
   & oracle  & 10(0)     & 0.06(0.08) & 0.94(0.08)  & 10(0) & 0.16(0.18) & 0.84(0.18) \\ 
   & first   & 1.5(1.5)  & 0(0.03)    & 0.14(0.14)  & 2.1(1.8) & 0.04(0.14) & 0.2(0.17) \\
10 & forward & 0.7(0.5)  & 0(0)       & 0.07(0.05)  & 0.8(0.4) & 0.01(0.11) & 0.08(0.04) \\
   & last    & 8.6(2)    & 0.03(0.06) & 0.84(0.19)  & 9.4(2.5) & 0.12(0.16) & 0.82(0.22) \\
   & RIC     & 9(2)      & 0.05(0.08) & 0.85(0.19)  & 6.2(3.5) & 0.09(0.15) & 0.57(0.33) \\
   & BIC     & 12(1.8)   & 0.17(0.11) & 0.98(0.05)  & 50(0) & 0.81(0.03) & 0.94(0.13) \\ 
  \hline
   & oracle  & 15(0)     & 0.07(0.07) & 0.93(0.07)  & 15(0) & 0.4(0.22) & 0.6(0.22) \\   
   & first   & 1.7(1.5)  & 0.01(0.09) & 0.11(0.1)   & 2.3(1.8) & 0.07(0.17) & 0.14(0.11) \\
15 & forward & 0.8(0.4)  & 0.01(0.1)  & 0.05(0.03)  & 0.8(0.4) & 0.04(0.2) & 0.05(0.03) \\
   & last    & 12.5(3.3) & 0.04(0.06) & 0.8(0.21)   & 11.4(4) & 0.29(0.22) & 0.53(0.25) \\
   & RIC     & 11.2(4.1) & 0.05(0.08) & 0.71(0.27)  & 3.6(2.8) & 0.11(0.2) & 0.21(0.17) \\
   & BIC     & 16.9(2.3) & 0.14(0.09) & 0.96(0.09)  & 50(0) & 0.78(0.07) & 0.73(0.23) \\ 

  \hline
\end{tabular}
\caption{Evaluation of model selection characteristics using several stopping rules.
  Here $n$ is fixed at 100, $p$ has 50 or 500 independent gaussian vectors,
  and the sparsity $k$ is set to 10 or 15.
  Nonzero coefficients have magnitudes in $[1.5\gamma, 2\gamma]$ for
  $\gamma = \sqrt{2\log (p)/n}$. The average over 400 simulations of the selected model
  size  (R), false discovery proportion  (FDP), and true positive proportion  (TPP)
  are shown with Monte Carlo standard errors in parentheses.}
\label{tab:easy}
\end{table}

\begin{table}[ht]
\centering
\begin{tabular}{|r|r|lll|lll|lll|}
 \hline
   \multicolumn{2}{|c|}{p} & \multicolumn{3}{c|}{50}  & \multicolumn{3}{c|}{500} \\ 
  \hline \hline
    k & rule  &  R        & FDP        &  TPP          &   R      & FDP        & TPP \\ 
  \hline
   & oracle  & 15(0)     & 0.18(0.09) & 0.82(0.09)  & 15(0) & 0.54(0.17) & 0.46(0.17) \\
   & first   & 1.1(1.1)  & 0(0.06)    & 0.07(0.07)  & 1.5(1.3) & 0.08(0.22) & 0.09(0.08) \\
15 & forward & 0.6(0.5)  & 0(0.05)    & 0.04(0.03)  & 0.7(0.4) & 0.05(0.21) & 0.05(0.03) \\
   & last    & 6.3(3.7)  & 0.03(0.07) & 0.4(0.23)   & 7.2(3.6) & 0.28(0.25) & 0.32(0.17) \\
   & RIC     & 6.5(3.1)  & 0.06(0.1)  & 0.41(0.2)   & 2.4(1.8) & 0.12(0.24) & 0.14(0.11) \\
   & BIC     & 14.2(2.9) & 0.16(0.1)  & 0.79(0.15)  & 50(0) & 0.84(0.05) & 0.54(0.18) \\
  \hline
   & oracle  & 20(0)     & 0.17(0.08) & 0.83(0.08)  & 20(0) & 0.65(0.12) & 0.35(0.12) \\
   & first   & 1.3(1.2)  & 0.02(0.11) & 0.06(0.06)  & 1.6(1.4) & 0.12(0.26) & 0.07(0.06) \\
20 & forward & 0.7(0.5)  & 0.02(0.13) & 0.03(0.02)  & 0.7(0.4) & 0.07(0.25) & 0.03(0.02) \\
   & last    & 7.7(4.3)  & 0.05(0.09) & 0.36(0.2)   & 8.8(4.2) & 0.41(0.23) & 0.24(0.11) \\
   & RIC     & 6.1(2.9)  & 0.05(0.11) & 0.29(0.14)  & 2(1.5) & 0.14(0.27) & 0.08(0.07) \\ 
   & BIC     & 17.2(3.9) & 0.14(0.09) & 0.74(0.16)  & 50(0) & 0.84(0.05) & 0.4(0.12) \\ 

  \hline
\end{tabular}
\caption{As in the previous table, but here the sparsity $k$ is set to 15 or 20
  and nonzero coefficients have magnitudes in $[1.1\gamma, 1.5\gamma]$ for
  $\gamma = \sqrt{2\log(p)/n}$. }
\label{tab:hard}
\end{table}

\section{Applications}
\label{sec:applications}
We now turn to applying forward stepwise and examining the behavior of
the $T\chi$ test in several unique settings, including a real data
example.

\subsection{Nonlinear regression with splines}

\begin{figure}[!htp]
\begin{center}
\subfigure[Power of forward stepwise with spline groups]{
\label{fig:spline:fwd}
\includegraphics[width=0.6\textwidth]{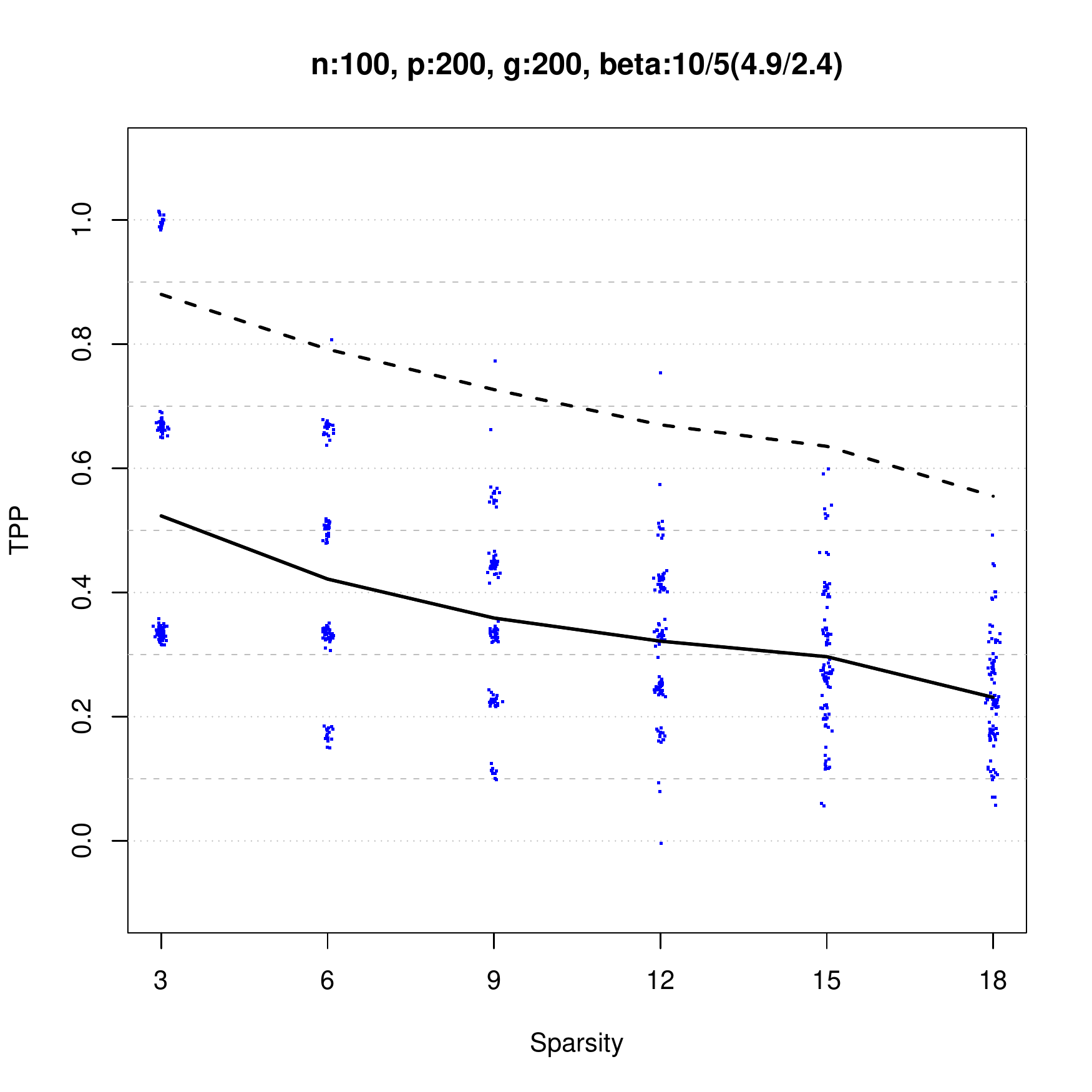}}
\hspace{-15pt}
\subfigure[$T\chi$ $p$-values with spline groups]{
\label{fig:spline:pval}
\includegraphics[width=0.6\textwidth]{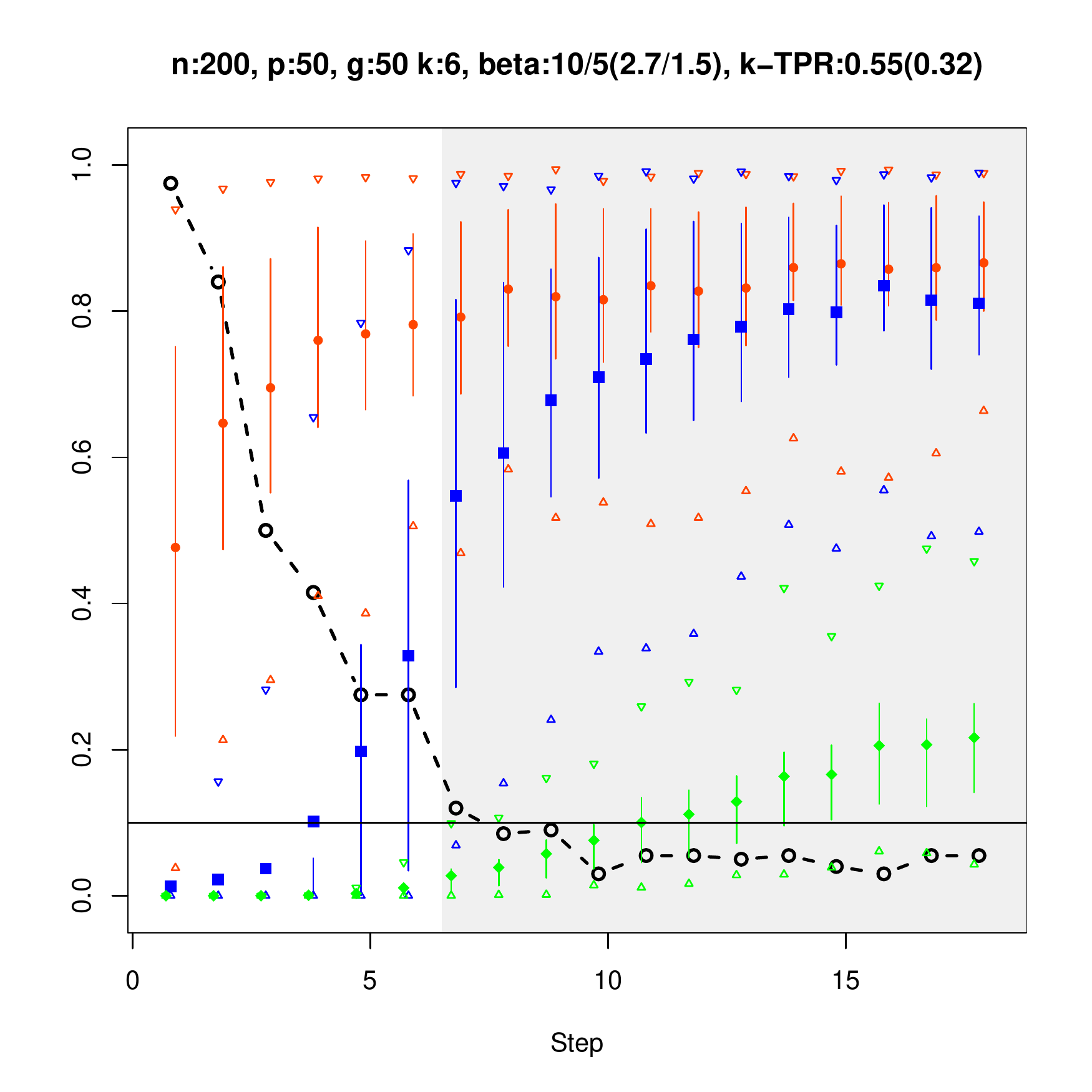}}
\caption{\small \it The first panel shows the power of forward stepwise
in this setting for various sparsity levels, with the dashed line
showing a liberal definition of power that allows mistakes when
choosing between a non-zero spline group and its corresponding linear
term.
The second panel shows the $T\chi$ $p$-value marginally
over each step, with red indicating the global null case, blue
indicating a signal with $k=6$ nonzero groups, and green showing
the nominal $\chi^2$ $p$-values also computed on the non-null case.}
\label{fig:spline}
\end{center}
\end{figure}

First we consider a simple extension of the usual linear regression setting to include nonlinear effects. Given a design matrix $X$, we augment this by adding additional columns given by nonlinear functions of the original covariates. Specifically, for each original covariate $X_g$ we compute a submatrix of spline basis vectors $X_g^s$. Adding these groups of spline basis vectors to the original design matrix yields a new design matrix
\begin{equation}
\label{eq:splinemat}
  \begin{aligned}
    \tilde X &=  \begin{pmatrix} X_1 & \cdots & X_G & X^s_1 & \cdots & X^s_G  \end{pmatrix}\\
  \end{aligned}
\end{equation}
with $\tilde G = 2G$ groups. Thus, at each step the procedure chooses between including a linear effect for each covariate or including a full spline basis if the true relationship is nonlinear.
For simplicity we assume the original groups are all size 1, and use the B-spline basis. In our simulations we generate original covariates uniformly in the interval [-1,1] and use cubic splines with boundary knots
\begin{equation}
  \begin{aligned}
    X_g &\sim \text{Unif}([-1,1]), \quad
    X_g^s = \text{CubicSplineBasis}(X_g).
  \end{aligned}
\end{equation}
Nonzero coefficients are split between original covariates and spline groups. Simulation results are shown in Figure~\ref{fig:spline}. Note that the varying group sizes present some difficulty to forward stepwise, which tends to first try linear approximations to nonzero spline groups before adding the true spline group. In Figure~\ref{fig:spline:pval} the average proportion of truly nonzero spline groups included in the first $k$ steps appears in parentheses next to the $k$-TPP (top right).

\subsection{Glinternet for hierarchical interactions}
\label{sec:glint}
In regression settings with many variables, choosing among models with pairwise interactions can drastically increase model complexity. \cite{glint} propose a method called \textsc{glinternet} to reduce the statistical complexity of this problem. The method imposes a strong hierarchical constraint on interactions (similar to that in \cite{bien:hierarchical}) where an interaction term can only be included if both its main effects are also included. They accomplish this by creating a design matrix with both main effects alone and also with groups including main effects with their first order interactions. Then they fit a group LASSO model with the expanded design matrix. Because interaction terms only appear in groups along with their respective main effects, the hierarchy condition holds for the fitted model. We now consider a related procedure as an example problem, but first modify their method to simplify some parts. Let the expanded design matrix be given by
\begin{equation}
\label{eq:glintmat}
\tilde X = \begin{pmatrix} X_1 & \cdots & X_G & X_{1:2} & \cdots & X_{1:G} & X_{2:3} & \cdots & X_{(G-1):G}  \end{pmatrix}
\end{equation}
where $X_{g:h}$ is the submatrix encoding the interaction between $X_g$
and $X_h$. For example, if both of these are categorical variables,
then $X_{g:h}$ consists of all
$p_gp_h$  column multiples between columns in group $g$ and columns
in group $h$. For example, if
\[
X_g = \begin{pmatrix} X_{g1} & X_{g2} \end{pmatrix}, \quad
X_h = \begin{pmatrix} X_{h1} & X_{h2} \end{pmatrix}
\]
are two categorical variables each with two levels, then
\[
X_{g:h} = \begin{pmatrix} X_{g1} * X_{h1} & X_{g1} * X_{h2} & X_{g1} * X_{h1} & X_{g2} * X_{h2} \end{pmatrix}
\]
where * denotes the pointwise product (Hadamard product) of vectors
(the $i$th entry of $X_{g1} * X_{h1}$ is the $i$th entry of $X_{g1}$ times
the $i$th entry of $X_{h1}$). For more details see \cite{glint}.
Note that this expanded matrix has
$\tilde G = G + \binom{G}{2} = O(G^2)$ groups. We refer to the first
$G$ of these as main effect groups and the remaining
as interaction groups. Finally, instead of fitting a
model by group LASSO, we use forward stepwise on the expanded design
matrix and call the resulting procedure FS-\textsc{glinternet}.
The overlapping groups still guarantee that our fitted model
satisfies the strong hierarchy condition. 

To demonstrate this method by simulation, we constructed signals which have
the first $k/3$ main effects nonzero but with no interactions, and the
remaining $2k/3$ nonzero main effects are matched to each other to form
interactions.
We also inflate each nonzero interaction
coefficient to be larger than the corresponding main
effect coefficients. This special case is favorable for our algorithm,
but our purpose here is merely to demonstrate the flexibility of the
hypothesis test and not to propose an optimal procedure for models with
interactions.

\begin{figure}[!htp]
\begin{center}
\subfigure[Power of FS-\textsc{glinternet} procedure]{
\label{fig:glint:fwd}
\includegraphics[width=0.6\textwidth]{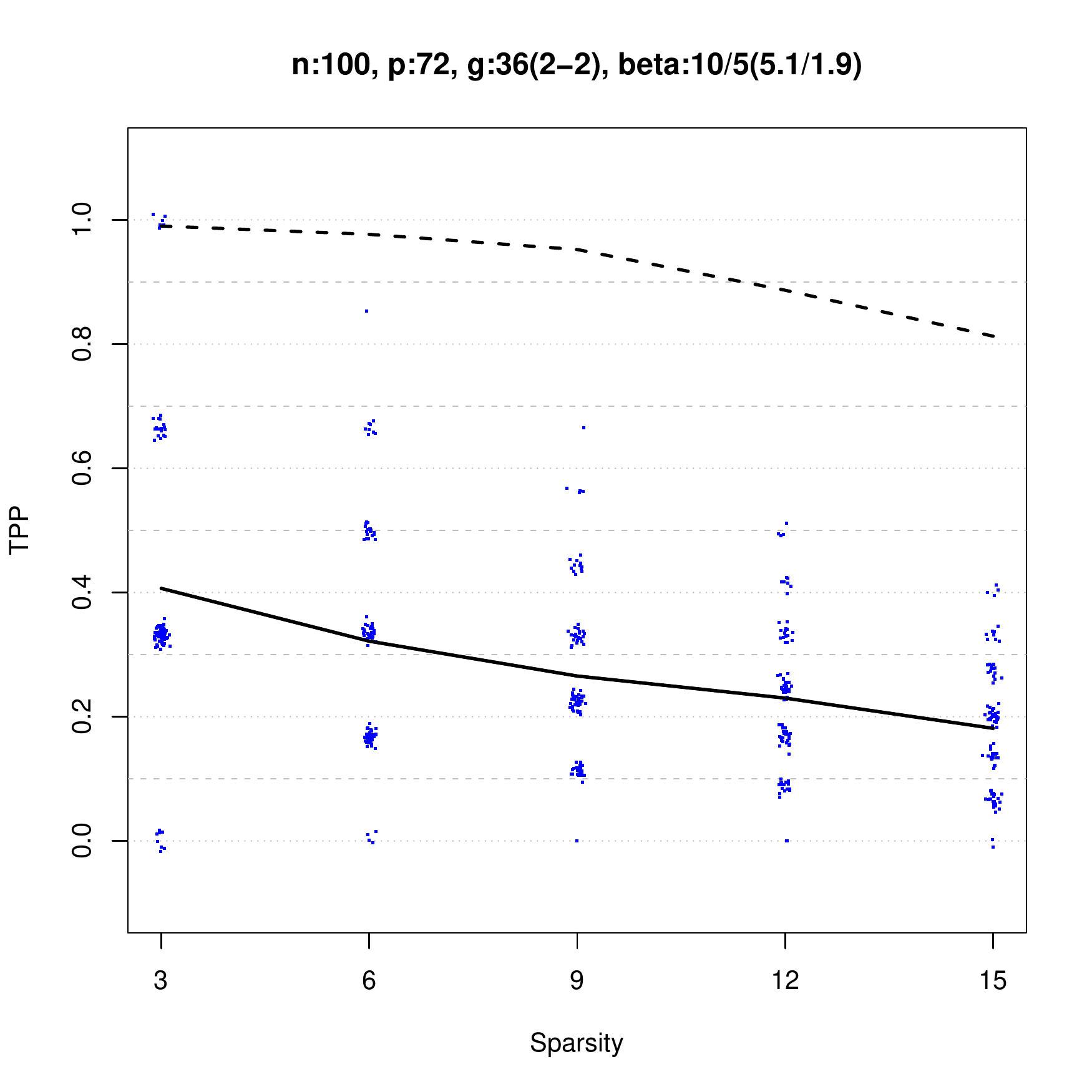}}
\hspace{-15pt}
\subfigure[$T\chi$ $p$-values with FS-\textsc{glinternet}]{
\label{fig:glint:pval}
\includegraphics[width=0.6\textwidth]{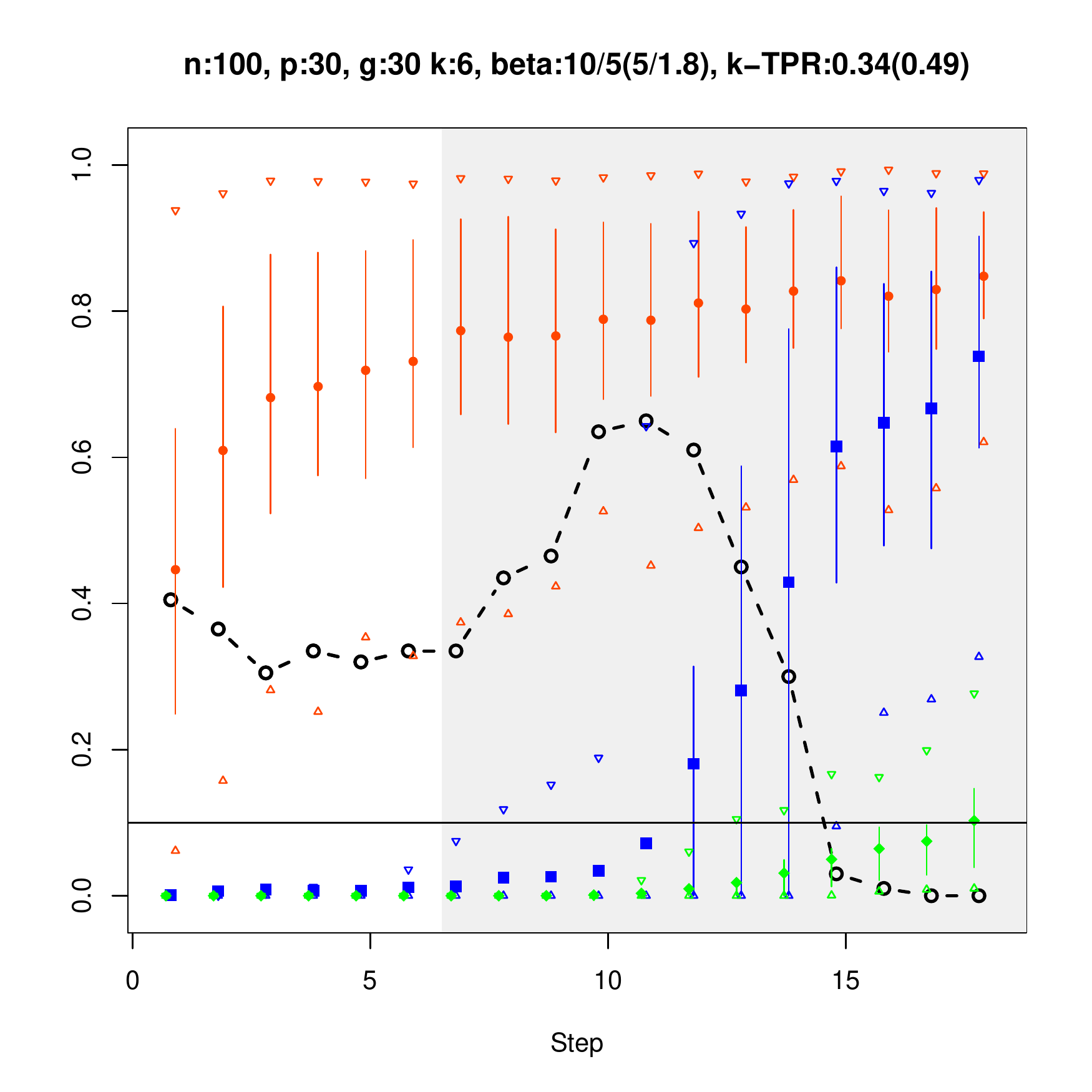}}
\caption{\small \it The first panel shows the power of forward stepwise
on the \textsc{glinternet} problem
for various sparsity levels. The dashed line shows a more forgiving
definition of power described in the section.
The second panel shows the $T\chi$ $p$-value marginally
over each step, with red indicating the global null case, blue
indicating a signal with $k=6$ nonzero groups, and green showing
the nominal $\chi^2$ $p$-values.}
\label{fig:glint}
\end{center}
\end{figure}

Results are shown in Figure~\ref{fig:glint}. The left
panel shows average power of forward stepwise. Power is calculated
using the group structure we impose, and not in terms of the original
main effects and interactions. However, the dashed line shows a more
forgiving definition of power where we are rewarded for discovering part
of a nonzero group, i.e. for discovering only one main effect from a
true interaction group. The proportion of nonzero interaction groups
that were discovered in the first $k$-steps is shown in parentheses
after the $k$-TPP (top right).

\subsection{HIVdb data example}
\label{sec:hiv}

\cite{HIV} use genomic information to predict efficacy of antiretroviral
drugs in treating HIV. Quantitative measurements of drug
response/resistance were regressed on categorical covariates encoding
the presence and type of genetic mutations in certain regions of the HIV
genome. We attempt a similar
analysis using forward stepwise, and report the $T\chi$ $p$-value at each step.
Categorical covariates are encoded as groups of dummy variables using
the full encoding, and these groups are normalized by their Frobenius
norm. We perform forward stepwise once
for each drug response, restricting to the subset of the data with
no missing values.

\begin{figure}[!htp]
\begin{center}
\includegraphics[width=0.95\textwidth]{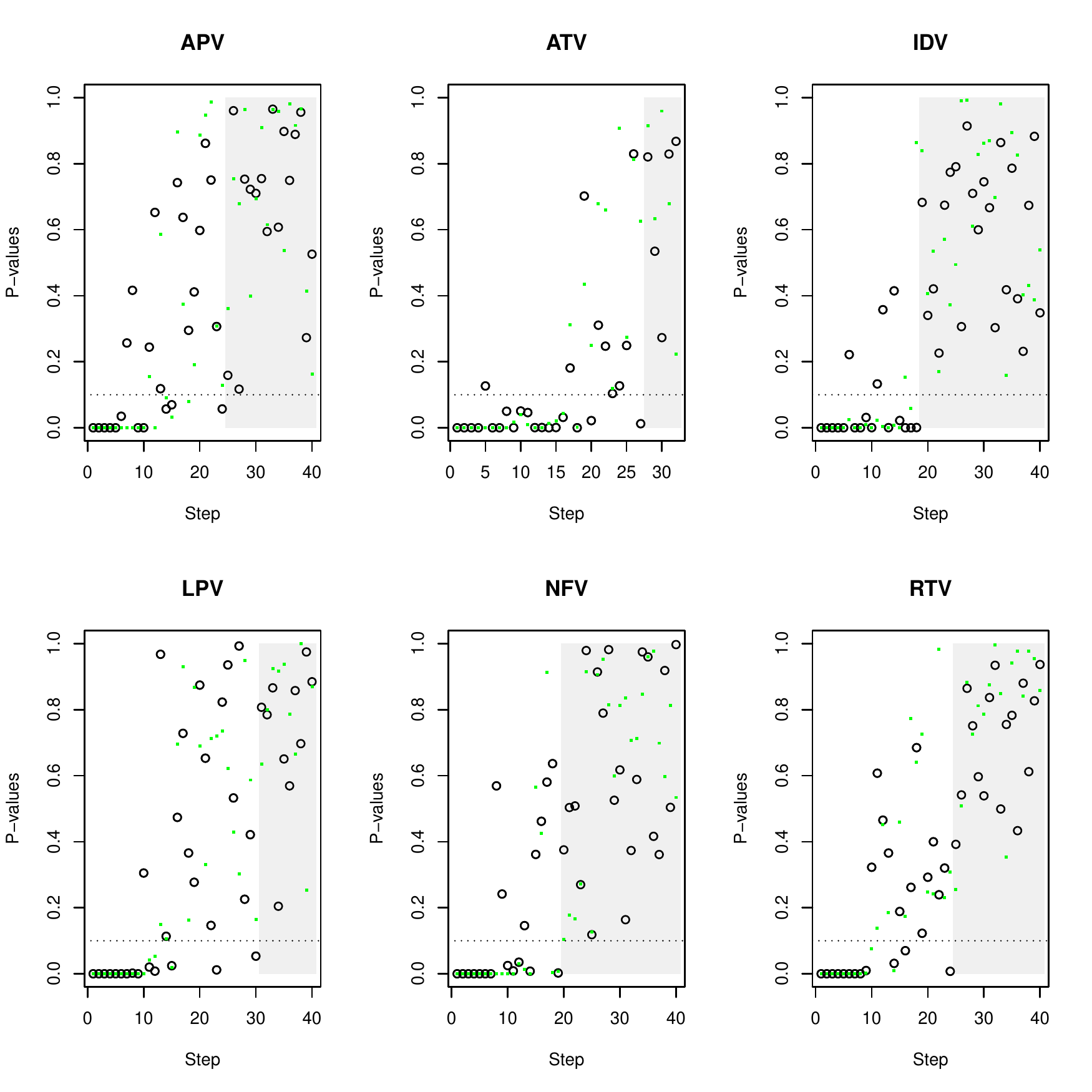}
\caption{\small \it Forward stepwise results from PI dataset}
\label{fig:HIVPI}
\end{center}
\end{figure}

\begin{figure}[!htp]
\begin{center}
\includegraphics[width=0.95\textwidth]{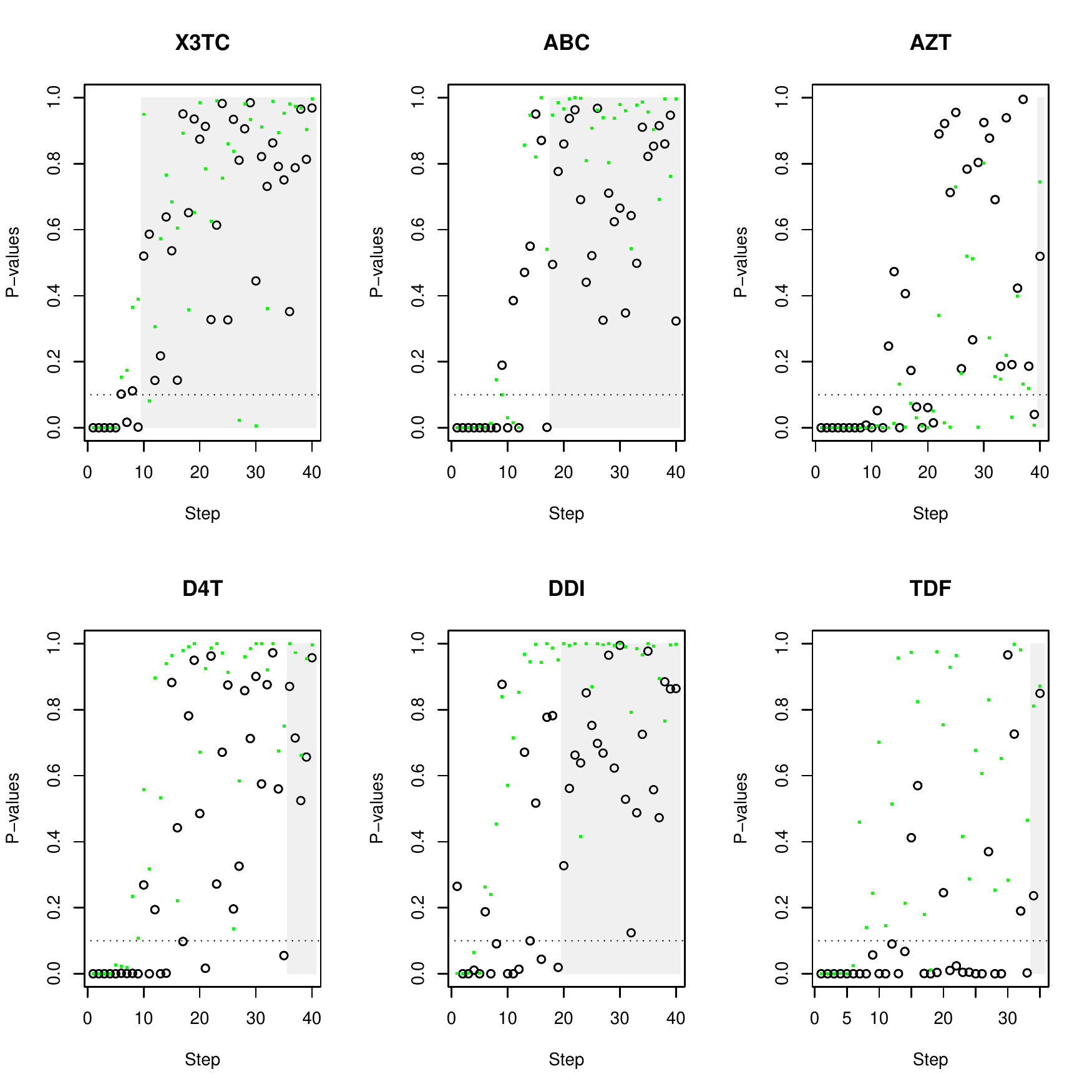}
\caption{\small \it Forward stepwise results from NRTI dataset}
\label{fig:HIVNRTI}
\end{center}
\end{figure}

Results from two data sets are displayed in Figure~\ref{fig:HIVPI} and
Figure~\ref{fig:HIVNRTI}. The PI data contains protease inhibitor
mutations, and the NRTI data contains nucleoside RT inhibitor mutations.
Each panel shows results for a different drug,
with $p$-values plotted by step in forward stepwise. 
The \textit{last} stopping rule \eqref{eq:klast} is applied and the region of the plot
corresponding to the chosen model is unshaded, with the remainder shaded.
The first several chosen variables are shown in Tables~\ref{tab:hivPI} and
\ref{tab:hivNRTI}.

\begin{table}[ht]
\centering
\begin{tabular}{rllll}
  \hline
 Drug & n & G & $\hat k$ & Selected variables \\ 
  \hline
 APV & 768 & 82 & 24 & P90 P46 P54 P84 P88 P32 P50 P76 P33 P10 P15 P82
\ldots \\ 
 ATV & 329 & 71 & 27 & P90 P54 P84 P50 P30 P32 P24 P76 P62 P46 P35 P88
 \ldots \\ 
 IDV & 827 & 82 & 18 & P90 P46 P54 P84 P82 P62 P88 P73 P35 P50 P71 P24
 \ldots \\ 
 LPV & 517 & 76 & 30 & P90 P54 P46 P84 P36 P82 P76 P47 P50 P10 P73 P33
 \ldots \\ 
 NFV & 844 & 82 & 19  & P90 P46 P30 P54 P84 P36 P88 P73 P24 P82 P50
 P71
 \ldots \\ 
 RTV & 795 & 82 & 24  & P90 P54 P46 P84 P82 P36 P24 P50 P32 P73 P13 P15
  \ldots \\ 
 SQV & 826 & 82 & 32 & P90 P84 P54 P30 P48 P36 P24 P53 P88 P73 P15 P82
 \ldots \\ 
   \hline
\end{tabular}
\caption{\em Variables chosen using \textit{last} stopping rule in forward
  stepwise on HIVdb PI dataset} 
\label{tab:hivPI}
\end{table}

\begin{table}[ht]
\centering
\begin{tabular}{rllll}
  \hline
Drug & n & G & $\hat k$ & Selected variables \\ 
  \hline
X3TC & 633 & 176 & 9 & P184 P41 P65 P67 P151 P210 P181 P83 P215 \\ 
ABC & 628 & 176 & 17 & P184 P41 P151 P67 P210 P65 P74 P83 P218 P215 P115 P69
\ldots  \\
AZT & 630 & 176 & 39 & P41 P67 P184 P151 P210 P70 P74 P215 P181 P77
P103 P69 \ldots \\ 
D4T & 630 & 176 & 35 &P41 P151 P67 P210 P184 P69 P65 P218 P215 P118 P75 P83
\ldots \\ 
DDI & 632 & 176 & 19 & P184 P151 P41 P74 P65 P67 P210 P218 P83 P75 P69
P118
\ldots \\ 
TDF & 353 & 153 & 33 &P41 P184 P70 P210 P65 P74 P181 P62 P215 P68 P67 P98
\ldots \\ 
   \hline
\end{tabular}
\caption{\em Variables chosen using \textit{last} stopping rule in forward
  stepwise on HIVdb NRTI dataset} 
\label{tab:hivNRTI}
\end{table}

\begin{figure}[!htp]
\begin{center}
\includegraphics[width=0.95\textwidth]{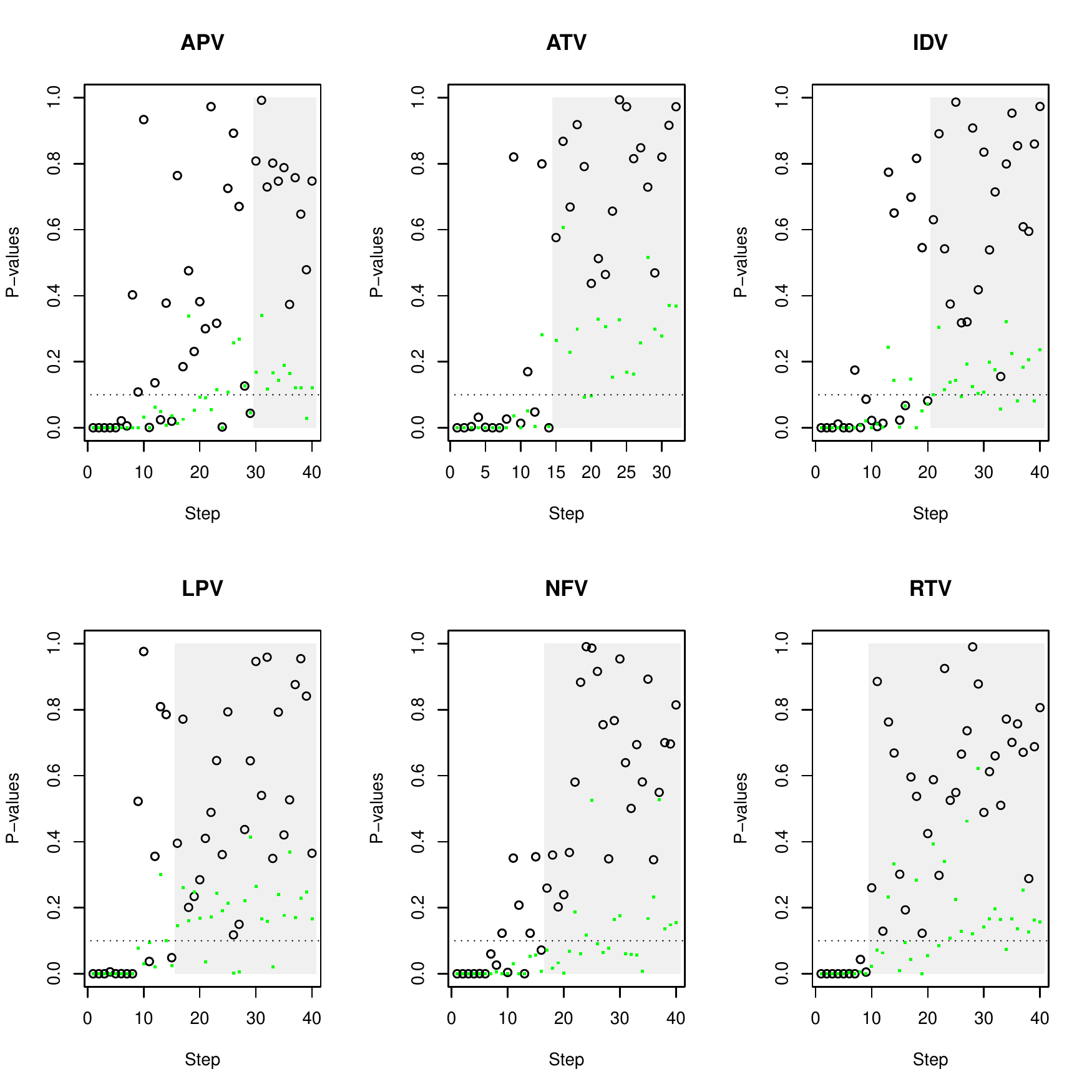}
\caption{\small \it \textsc{Glinternet} results from PI dataset}
\label{fig:HIVPI:glint}
\end{center}
\end{figure}

\begin{figure}[!htp]
\begin{center}
\includegraphics[width=0.95\textwidth]{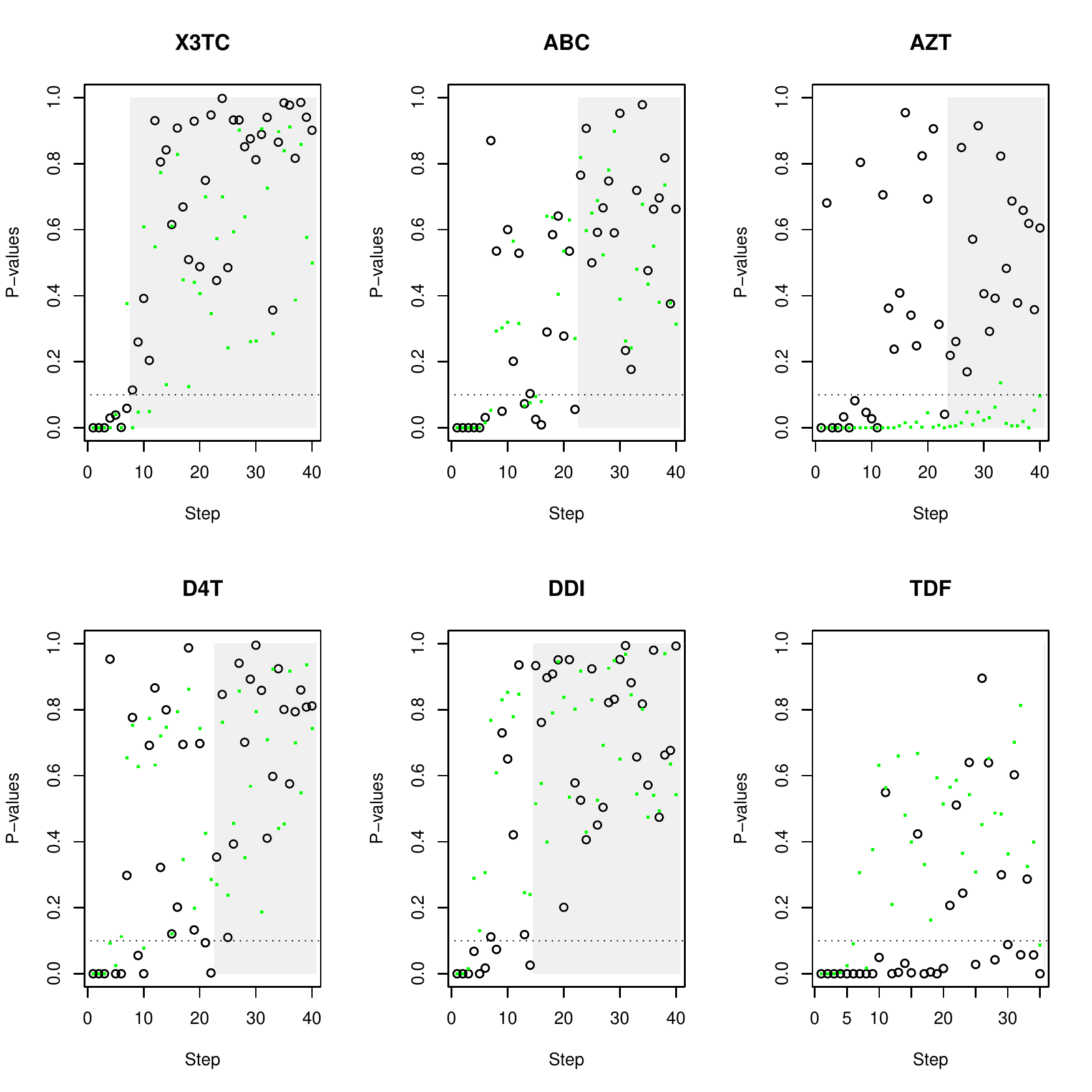}
\caption{\small \it \textsc{Glinternet} results from NRTI dataset}
\label{fig:HIVNRTI:glint}
\end{center}
\end{figure}

We can also use the \textsc{Glinternet} procedure described in Section~\ref{sec:glint} to fit a model with pairwise interactions. Results from this are shown in Figures~\ref{fig:HIVPI:glint} and \ref{fig:HIVNRTI:glint} and Tables~\ref{tab:glintPI} and \ref{tab:glintNRTI}.

\begin{table}[ht]
\centering
\begin{tabular}{rllll}
  \hline
Drug & n & G & $\hat k$ & Selected variables \\ 
  \hline
APV & 768 & 3403 & 29 &P8*P10 P7*P46 P54*P84 P76*P90 P30*P88 P33*P50
\ldots \\
ATV & 329 & 2556 & 14&P10*P76 P8*P71 P20*P46 P33*P48 P50*P84 P79*P90 \ldots \\
IDV & 827 & 3403 & 20&P10*P26 P46*P54 P48*P90 P20*P24 P82*P84 P71*P88 \ldots \\
LPV & 517 & 2926 & 15&P54 P46*P48 P10*P75 P20*P84 P82*P84 P33*P50 \ldots \\
NFV & 844 & 3403 & 16&P10*P30 P46*P90 P54*P88 P20*P84 P82*P84 P48*P73 \ldots \\
RTV & 795 & 3403 & 9&P82*P84 P24*P90 P39*P54 P36*P46 P10*P88 P63*P93 \ldots \\
SQV & 826 & 3403 & 13&P10*P76 P84*P90 P54*P88 P36*P48 P73*P74 P71*P93 \ldots \\
   \hline
\end{tabular}
\caption{\em Variables chosen using \textit{last} stopping rule in
  \textsc{Glinternet} on HIVdb PI dataset} 
\label{tab:glintPI}
\end{table}

\begin{table}[ht]
\centering
\begin{tabular}{rllll}
  \hline
  Drug & n & G & $\hat k$& Selected variables \\ 
  \hline
X3TC & 633 & 15576 & 7& P157*P184 P65*P215 P184*P215 P151*P210 P67*P75 \ldots \\ 
ABC & 628 & 15576 & 22&P54*P184 P151*P215 P65*P210 P69*P74 P184*P215 \ldots \\
AZT & 630 & 15576 & 23&P77*P215 P67*P151 P151*P184 P70*P210 P135*P214 \ldots \\
D4T & 630 & 15576 & 22&P151*P215 P65*P210 P67*P69 P104*P184 P75*P218 \ldots \\
DDI & 632 & 15576 & 14&P41*P151 P65*P184 P69*P74 P62*P219 P75*P210 \ldots \\
TDF & 353 & 11781 & 35&P41*P65 P184*P224 P32*P70 P68*P210 P174*P181 \ldots \\
   \hline
\end{tabular}
\caption{\em Variables chosen using \textit{last} stopping rule in
  \textsc{Glinternet} on HIVdb NRTI dataset} 
\label{tab:glintNRTI}
\end{table}

\section{Discussion}
\label{sec:discuss}

Under the global null hypothesis $\beta = 0$, we have a test statistic
which we can use against the alternative of including the ``best''
predictor (the one chosen by forward stepwise).
This test statistic has an exact, finite sample distribution.
If we choose to include the variable, we are no longer in the global null
setting. However, by orthogonalizing both the response and the remaining
predictors with respect to the included variable, it is reasonable to
iterate the global null test. While not fully theoretically justified,
this method seems to work well in all our simulations. By this we mean that
the $T\chi$ $p$-value tends to be small on the step when including the last truly
nonzero predictor, uniform on the following step, and subsequently
stochastically larger than uniform. This is in contrast to $p$-values
calculated from traditional variable-inclusion tests like the $\chi^2$
test, which tend to be smaller than uniform long after the last truly
nonzero predictor has been included.

When adding the next variable in forward stepwise, our hypothesis test
roughly depends on the improvement gained by this variable compared to
the next best variable. Thus, the $T\chi$ $p$-value is small when there is a large
gap between the variable to be included and all the remaining variables.
If multiple predictors have truly nonzero coefficients that are close
in magnitude, the $p$-value may be large until forward stepwise reaches
the last one. This motivated us to consider the \textit{last} stopping
rule for model selection \eqref{eq:klast}. In simulations we
found this stopping rule to have good performance in terms of power
(true positive rate), comparable to that of RIC \citep{RIC}.
But it does not control the false discovery rate
unless the truly nonzero coefficients are large enough to guarantee
forward stepwise picks the corresponding variables before picking too
many noise variables.

The present work calculates $p$-values at each step ignoring the constraints
imposed by previous steps. Future work adjusting for all previous steps
is in progress, and may be able to give exact $p$-values at a step chosen
stochastically by procedures like BIC and RIC. The authors will
release an R package implementing these methods.

\textbf{Acknowledgements:}
The authors would like to thank Robert Tibshirani and Trevor Hastie for many helpful comments and suggestions.

\bibliographystyle{my-agsm}
\bibliography{paper}

\appendix
\section{Alternate derivation and algorithm for $T\chi$ test}
\label{app:algo}

\begin{algorithm}
 \caption{Computing the $T\chi$ $p$-value}
 \label{algo:pval}
 \begin{algorithmic}
   \REQUIRE Response $y$, grouped design matrix $X$ with weights, inactive set $A^c$, index $g$ of last group to enter active set $A$.
   \ENSURE $T\chi$ $p$-value for the group $g$ entering the model.
   \STATE Compute $H_g$ and $\sigma^2_g$
   \IF{$p_g = 1$}
   \STATE $\sigma_g^2 \gets X_g^T\Sigma X_g/w_g^2$
   \STATE $\tilde X_g \gets X_g / w_g \cdot \text{sign}(X_g^Ty)$
   \ELSE
   \STATE $P_g \gets \Sigma XV_g (V_g^T X^T \Sigma X V_g)^\dagger V_g^TX^T$
   \STATE $H_g \gets (I - P_g)\Sigma$
   \STATE $\sigma^2_g \gets y^TX_gX_g^T H_g X_gX_g^Ty / (w_g^2 \norm{X_g^Ty}_2^2)$
   \STATE $\tilde X_g \gets H_g X_g X_g^T y /(\norm{X_g^Ty}_2 w_g)$
   \ENDIF
   \STATE $\lambda \gets \norm{X_g^Ty}_2/w_g$
   \STATE \COMMENT{Compute the following two $p$-vectors}
   \STATE $a \gets X^T(y - \lambda \tilde X_g)$
   \STATE $b \gets X^T \tilde X_g$
   \STATE Compute solution $(v_-, v_+)$ of \textsc{LinearFractional}(a,b) optimization subproblem
   \STATE $r \gets \text{rank}(X_g)$
   \STATE $u \gets [F_{\chi^2_r}(v^2_-/\sigma_g^2) - F_{\chi^2_r}(\lambda^2/\sigma_g^2)]/[F_{\chi^2_r}(v^2_-/\sigma_g^2) - F_{\chi^2_r}(v^2_+/\sigma_g^2)]$
   \RETURN $u$
 \end{algorithmic}
\end{algorithm}

In this section we describe an alternative derivation following the
discussion of group LASSO in \cite{tests:adaptive}. The implementation
described here was used in all simulations.
For the rest of this section let $g = \gstar$ be the index of
the group attaining the maximum on line 3 of Algorithm~\ref{algo:fs}.
For a vector $u \in \real^p$, let $u_h$ denote the coordinates
of $u$ corresponding to the columns of group $h$ in the design matrix $X$.
We can rearrange the columns of $X$ to group these adjacently, so that
\[
u^TX^T = \begin{pmatrix} u_1^T X_1^T & u_2^T X_2^T & \cdots & u_G^TX_G^T \end{pmatrix}
\]
One step of the calculation will be to find an orthonormal basis for the
linear space $\vecsp_g = \{ u \in \real^p : u_g^T X_g^T y = 0, u_h = 0
\text{ for all } h \neq g \}$ so that we can project orthogonally to
this space. If $X_g$ is a single column and $X_g^Ty \neq 0$ (which
should be the case since $g$ maximizes the absolute value of this
quantity), then the
space is trivial and the desired orthogonal projection is the identity.
Otherwise, if $X_g$ has $p_g > 1$ columns the space $\vecsp_g$
generally has dimension $p_g-1$.

We compute an orthonormal basis by Gram-Schmidt and form a
$p_g \times p_g$ matrix, which we denote $V_g$, by appending
0's in the coordinates corresponding to all groups $h \neq g$ and
an additional column of zeroes (since Gram-Schmidt only produces $p_g-1$).
We can now define the projection

\begin{equation}
 \begin{aligned}
   \label{eq:proj}
   P_g &= \Sigma X_gV_g (V_g^T X_g^T \Sigma X_g V_g)^\dagger V_g^TX_g^T. \\
  \end{aligned}
\end{equation}
Also define $H_g = (I-P_g)\Sigma$ and the conditional variance

\begin{equation}
 \begin{aligned}
   \label{eq:cvar}
   \sigma^2_g &= y^TX_gX_g^T H_g X_gX_g^Ty / (w_g^2 \norm{X_g^Ty}_2^2) \\
 \end{aligned}
\end{equation}
These simplify when $X_g$ is a single column, in which case $P_g = 0$,
$H_g = \Sigma$, and $\sigma^2_g = X_g^T\Sigma X_g^T/w_g^2$.
Note that $F_{\chi^2_r}$ denotes the distribution function of a $\chi^2$
random variable with $r$ degrees of freedom.

Next we describe the \textsc{LinearFractional} optimization
subproblem and its solution.
The problem was named as it originated in the form
\[
\maximize_{h \neq g, \norm{u_h}_2 = 1}
\frac{u_h^TX_h^Ty - u_h^TX_h^T\tilde X_g \tilde X_g^Ty}
{1 - u_g^T \tilde X_g^TX_hu_h}
\]
The solution we describe next is to a slightly different problem which
also incorporates the information that $g$ maximizes $\norm{X_g^Ty}_2/w_g$.
Although the logic seems a bit complicated, it mainly involves ruling out
several cases as infeasible. The infeasibilities are precisely those given
by the characterization of the global maximizer in \cite{tests:adaptive}.
After ruling out degenerate cases, we obtain the following solution
by transforming to trigonometric coordinates and using calculus.

\begin{algorithm}
 \caption{The \textsc{LinearFractional} subproblem}
 \label{algo:linfrac}
 \begin{algorithmic}
   \REQUIRE The $p$-vectors $a, b$ in Algorithm~\ref{algo:pval}, weights, inactive set $A^c$, a small tolerance number
(we take $1e^{-10}$)
   \ENSURE Solution pair $(v_-, v_+)$.
   \FOR{$h$ in $A^c$}
     \IF{$\norm{b_h}_2 == 0$ or $\norm{a_h}_2/\norm{b_h}_2 < \text{tol}$}
       \STATE $(v^-_h, v^+_h) \gets (0, \infty)$
     \ELSE
       \STATE $\theta_c \gets a_h^Tb_h/(\norm{a_h}_2 \norm{b_h}_2)$
       \STATE $\theta_s \gets \sqrt{1-\theta_c^2}$
       \STATE $\theta \gets \arccos (\theta_c)$
       \STATE $\phi_s \gets \theta_s \norm{b_h}_2 / w_h$
       \IF{$\phi_s > 1$}
         \STATE $(v^-_h, v^+_h) \gets (0, \infty)$
       \ELSE
         \STATE $\phi \gets \arcsin(\phi_s)$
         \STATE $\phi_2 \gets \pi - \phi_1$
         \STATE $z_\pm \gets s_\pm \norm{a_h}_2 \cos(\phi) /(w_h - s_\pm \norm{b_h}_2 \cos(\theta - \phi))$ for $s_\pm = \pm 1$
         \IF{$\norm{b_h}_2 < w_h$}
           \STATE $(v_h^-, v_h^+) \gets (\max \{ z_+, z_- \}, \infty)$
         \ELSE
           \STATE $(v_h^-, v_h^+) \gets (\min \{ z_+, z_- \}, \max \{ z_+, z_- \})$
         \ENDIF
       \ENDIF
     \ENDIF
   \ENDFOR
   \STATE $v_- \gets \max_h v_h^-$
   \STATE $v_+ \gets \min_h v_h^+$
   \RETURN $(v_-, v_+)$
 \end{algorithmic}
\end{algorithm}

\section{A note on power}
\label{app:power}

Although we have shown through simulations in a variety of settings that
the $T\chi$ test has power, here we give a theoretical result in one specific case
where it is easy to prove something. Namely, consider the case where all
groups have size 1, and the design matrix is orthonormal. By a
transformation we can reduce to the identity design case where $n=p$,
$X=I$, so
\[
y_i = \mu_i + \epsilon_i, \qquad \epsilon_i \sim N(0,1).
\]
In this simple case the $T\chi$ $p$-value can be calculated as
\[
\frac{\tilde \Phi(\lambda_1)}{\tilde \Phi(\lambda_2)}
\]
where $\lambda_1$ and $\lambda_2$ are the first two knots in the LASSO
solution path and $\tilde \Phi = 1 - \Phi$ is the survival function of
a standard Gaussian.

The simplest possible alternative hypothesis
is the 1-sparse case where a single $\mu_i$ is nonzero. Let
$\mu_1 = r \sqrt{2\log(p)}$ and all other $\mu_i = 0$. If $r$ is some
constant greater than 1, then with high probability the first
knot $\lambda_1$ will be achieved by $y_1$. 
Applying the normal tail bounds
\[
\frac{\tilde \Phi(xr)}
{\tilde\Phi(x)} \leq \frac{\phi(rx)/rx}{(1-1/x^2)\phi(x)/x}
\]
with $x = \sqrt{2\log(p)}$ we obtain, by continuity,
\[
 \frac{\tilde \Phi(\lambda_1)}{\tilde \Phi(\lambda_2)}
\lessapprox \frac{p^{1-r^2}}{r\left(1-\frac{1}{2\log(p)}\right)}.
\]
For any $r > 1$ the upper bound goes to 0 as $p \to \infty$, so
in this case the test has asymptotic full power at the same threshold
as Bonferroni. This is the best possible threshold for asymptotic power
against the 1-sparse alternative.

\section{A note on parallel computation}
\label{app:computation}

One strength of forward stepwise is that it can be computed in parallel, hence can handle situations with many groups of predictor variables. We now describe, at very a high level, how to accomplish this. Suppose $m > 1$ machines are available for use. Partition the set of predictor variable groups, along with their respective weights, into $m$ disjoint blocks and store each block on its own machine. Now at each step of Algorithm~\ref{algo:fs}, the residual $r_{s-1}$ is broadcast to all machines, and each machine computes $\| X_g^T r_{s-1} \|_2/w_g$ for all the groups $g$ stored on that machine. Each machine reports its own maximum and maximizer $g^*$, and the global maximizer is found. The projector matrix $P_{g^*}$ is computed and broadcast to all machines which in turn use it to project all their predictors.

We can also verify that Algorithms \ref{algo:pval} and \ref{algo:linfrac} only require block matrix operations with blocks given by individual groups
or subproblems computable separately for each group. Hence, the $T\chi$ $p$-value can also be computed in parallel this way.

\end{document}